\documentclass[12pt]{iopart}
\usepackage{graphicx}

\begin{document}
\title[Entanglement Dynamics of Kicked Qubits]{Control and manipulation of entanglement between two coupled qubits by fast pulses} 
\author{Ferdi Altintas and Resul Eryigit}
\address{Department of Physics, Abant Izzet Baysal University, Bolu, 14280-Turkey.}
\ead{altintas$\textunderscore$f@ibu.edu.tr, resul@ibu.edu.tr}
\begin{abstract}
We have investigated the analytical and numerical dynamics of entanglement for two qubits that interact with each other via Heisenberg XXX-type interaction and subject to local time-specific external kick and Gaussian pulse-type magnetic fields in $x$-$y$ plane. The qubits have been assumed to be initially prepared in different pure separable and maximally entangled states and the effect of the strength and the direction of external fast pulses on concurrence has been investigated. The carefully designed kick or pulse sequences are found to enable one to obtain constant long-lasting entanglement with desired magnitude. Moreover, the time ordering effects are found to be important in the creation and manipulation of entanglement by external fields. 
\end{abstract}

\maketitle
\section{Introduction}
Entanglement exhibited by composite correlated quantum systems has become one of the most widely investigated physics subject of recent years~\cite{nc}. It provides promising resource for some quantum information tasks, such as quantum teleportation~\cite{bbcjpw}, cryptography~\cite{ekert} and computing~\cite{gruska}. For those applications, the generation, manipulation, detection and control of entanglement between the quantum systems should be precise and effectively done during carrying out of the required operations. A spin $1/2$ particle in a magnetic field is found to be a suitable candidate as a qubit of a quantum computer~\cite{dldpd}, so the Heisenberg chains have been used to construct a quantum computer based on quantum dots~\cite{dldpd}, nuclear spins~\cite{kane}, electronic spins~\cite{vrijen} and optical lattices~\cite{sorensen}. Moreover, Wu {\it et al.} showed that one qubit gates in spin-based quantum computers can be constructed with a global magnetic field and controllable Heisenberg exchange interactions~\cite{wl}.  Imamoglu {\it et. al.} demonstrated that 1-D Heisenberg chain of spin 1/2 particles~(qubits) can be used as a prototypical system to study the role of entanglement in quantum computational tasks~\cite{iabdlss}.  

Numerous studies have been devoted to the control, production and manipulation of entanglement in Heisenberg spin chains with the help of external magnetic fields~\cite{ms,sla,wbsb,model2,wang,sb,zhsk,zhsk2,bd,fare,ask}. Among them, Sadiek {\it et al.} investigated the time evolution of entanglement between two Heisenberg XYZ coupled qubits under an external sinusoidal time-dependent nonuniform magnetic field and they demonstrated that the entanglement of the system can be controlled and tuned by varying the time-dependent magnetic field and the Heisenberg exchange coupling parameters~\cite{sla}. Abliz {\it et al.} studied the entanglement dynamics of two-qubit Heisenberg XYZ model affected by population relaxation and subject to various types of external magnetic fields, such as an inhomogeneous static field, homogeneous exponentially varying and periodically varying magnetic fields, and they demonstrated that high entanglement can be produced, controlled and modulated with the help of external time-varying fields despite the existence of dissipation~\cite{model2}. Huang and Kais demonstrated that the entanglement in one-dimensional spin system, modeled by the XY Hamiltonian can be localized between nearest-neighbor qubits for certain values of the step function external magnetic fields~\cite{zhsk}. In Ref.~\cite{ms},  the possibility of creating and controlling entangled states by changing the relative phase of control pulses was investigated. Huang {\it et al.} demonstrated that the dynamics of the nearest-neighbor entanglement for one-dimensional spin system under sinusoidal magnetic fields displays a periodic structure with a period related to that of the magnetic field~\cite{zhsk2}. Blaauboer and Di Vincenzo presented a scheme for detecting entanglement between electron spins in a double-quantum-dot nanostructure and they demonstrated separation and coherently rotating of entangled spins in quantum dots by using a time-dependent gate voltage and magnetic field~\cite{bd}. 

The form of the external magnetic field used in control of entanglement is an important parameter in manipulation of the entanglement between the qubits. As summarized above, the most of the external fields studied for the control are in static, exponential, step or sinusoidal form. An external field of the kick or Gaussian pulse sequence form might provide a better control schema for entanglement. Along these lines Kaplan {\it et al.} and Shakov {\it et al.} have investigated the population and coherence dynamics of a single qubit under the influence of kick or Gaussian pulse sequence magnetic fields~\cite{kaplan,shakov}. It was shown that the instantaneous pulses~(kicks) provide a full population transfer in a qubit from one state to the other which has a great deal of interest in quantum computing and control theory~\cite{wjcan}. In our recent paper~\cite{fare}, we have extended the schema introduced in Refs.~\cite{kaplan,shakov} to two coupled qubits and have investigated the dynamics of entanglement  between two Heisenberg qubits under the influence of strong delta function~(kick) and Gaussian pulse type magnetic fields directed along the $z$-axes. The qubits have been assumed to be  initially prepared in the separable $\left|01\right\rangle$ and the maximally entangled $\frac{1}{\sqrt{2}}(\left|10\right\rangle+\left|01\right\rangle)$ states. We have shown that the fast pulses provide an efficient way of controlling entanglement between the coupled qubits and also by this control it is possible to manipulate the transition from disentangled to the entangled states of the system. Longitudinal control fields were found to be ineffective in creating entanglement for the $\left|11\right\rangle$ and $\left|00\right\rangle$ type initial states or manipulating the entanglement of $\frac{1}{\sqrt{2}}(\left|11\right\rangle\pm\left|00\right\rangle)$ type maximally entangled initial states.

In the present study, we extend the work in Ref.~\cite{fare} to consider the effect of the transverse field and analyze the dynamics of entanglement between two qubits that interact with each other via Heisenberg XXX type interaction that are subject to site-specific time-dependent magnetic control fields. Our aim is to widen the possibility of creating and manipulating entanglement with the tailored external kick and Gaussian pulse sequence type fields for arbitrary pure initial states. One important finding of the present work is the transverse external fields can control the dynamics of entanglement for two qubits for initial states that could not be manipulated with the longitudinal fields. We have also showed that the opposite scenario can take place for transverse external fast pulses.

The paper is organized as follows. In Sec.~\ref{secbasic}, the model and basic formulation for the solution of time evolution are briefly discussed. In Sec.~\ref{concurrence}, Wootters concurrence as a measure of entanglement is introduced. Entanglement dynamics of two qubits under  multiple kicks analytically and multiple Gaussian pulses numerically are investigated in Secs.~\ref{seckicked} and~\ref{gaussian}, respectively. In Sec.~\ref{conclusion}, we conclude with a summary of important results.

\section{The model and basic formulation}
\label{secbasic}
In the present study, we consider two Heisenberg XXX-coupled qubits under time-dependent external magnetic fields in $x$-$y$ plane. The time-dependent Hamiltonian for this model can be represented as~\cite{model2,model1}~($\hbar=1$):
\begin{eqnarray}
\label{totalH}
\hat{H}(t)=\hat{H}_0+\hat{H}_{int}(t),
\end{eqnarray}
where
\begin{eqnarray}
\label{individualH}
\hat{H}_0&=&J\displaystyle\sum_{i=x,y,z}\hat{\sigma}^{i}_1\hat{\sigma}^{i}_2,\nonumber\\
\hat{H}_{int}(t)&=&\frac{1}{2}\displaystyle\sum_{i=1}^2B_i(t)(\cos(\theta)\hat{\sigma}^x_i+\sin(\theta)\hat{\sigma}^y_i),
\end{eqnarray}
where $\hat{\sigma}^i_{1,2}~(i=x,y,z)$ are the usual Pauli spin operators, $B_{1}(t)$ and $B_{2}(t)$ are the external time-dependent magnetic fields acting on the qubits $1$ and $2$, respectively, $J$ is the qubit-qubit interaction strength and $\theta$ is the angle between the magnetic fields and the $x$-axes. For simplicity, we will assume $0\leq\theta\leq\frac{\pi}{2}$. Here, we take $J$ to be constant in time and assume all the time dependence in the systems' Hamiltonian $\hat{H}(t)$ comes only from $\hat{H}_{int}(t)$. It is also possible to control qubit-qubit entanglement evolution by time dependent coupling strength instead of magnetic fields~\cite{lcms}. It was shown that such controls can be implemented physically, for example by using interacting flux qubits~\cite{lcms}.

The most general form of an initial pure state of the two-qubit system may be given by the state vector $\left|\Psi(0)\right\rangle=a_1(0)\left|11 \right\rangle+a_2(0)\left|10\right\rangle+a_3(0)\left|01\right\rangle+a_4(0)\left|00\right\rangle$, where $a_i(0)~(i=1,2,3,4)$ are complex numbers with $\displaystyle\sum_{i=1}^4|a_i(0)|^2=1$. Then, under Hamiltonian~(\ref{totalH}) the probability amplitudes evolve in time according to Schr\"{o}dinger equation as:
\begin{equation}
\label{hamiltonian}
i{d \over dt}\left [\begin{array}{c} a_1(t) \\ a_2(t) \\ a_3(t) \\ a_4(t)\end{array}\right]=\left[\begin{array}{cccc} J & \widetilde{B}_2(t)& \widetilde{B}_1(t) & 0 \\ \widetilde{B}_2(t)^* & -J & 2 J & \widetilde{B}_1(t) \\ \widetilde{B}_1(t)^* & 2 J & -J & \widetilde{B}_2(t) \\ 0 & \widetilde{B}_1(t)^* & \widetilde{B}_2(t)^*& J \end{array} \right ]
\left [\begin{array}{c} a_1(t) \\ a_2(t) \\ a_3(t) \\ a_4(t) \end{array} \right],
\end{equation}
where $\widetilde{B}_k(t)=\frac{1}{2}e^{-i \theta}B_k(t)~(k=1,2)$ and $\widetilde{B}_k(t)^*$ is its complex conjugate. Mathematically and conceptually, it is convenient to write the formal solution of Eq.~(\ref{hamiltonian}) in terms of the time evolution matrix $\hat{U}(t)$ as:
\begin{eqnarray}
\label{amps}
\left[\begin{array}{c} a_1(t) \\ a_2(t) \\ a_3(t) \\ a_4(t)\end{array}\right]=\hat{U}(t)
\left[\begin{array}{c} a_1(0) \\ a_2(0) \\ a_3(0) \\ a_4(0)\end{array}\right].
\end{eqnarray}
Here all the time-dependence of the system is contained in the time evolution operator $\hat{U}(t)$, while the initial conditions are specified in $\left|\Psi(0)\right\rangle$. The time evolution operator $\hat{U}(t)$ may be expressed as:
\begin{eqnarray}
\label{U}
\hat{U}(t)&=&\stackrel{\leftarrow}{T}e^{-i\int^t_0\hat{H}(t')dt'}=\stackrel{\leftarrow}{T}e^{-i\int^t_0\left(\hat{H}_0+\hat{H}_{int}(t')\right) dt'}\nonumber\\
&=&\stackrel{\leftarrow}{T}\sum_{n=0}^\infty\frac{(-i)^n}{n!} \int_0^t\hat{H}(t_n)dt_n ...\int_0^t\hat{H}(t_2)dt_2\int_0^t\hat{H}(t_1)dt_1.
\end{eqnarray}
Here $\stackrel{\leftarrow}{T}$ is the so-called the Dyson time ordering operator which arranges the operators in order of increasing of time~\cite{dysont}, for example, $\stackrel{\leftarrow}{T}\hat{H}(t_i)\hat{H}(t_j)=\hat{H}(t_j)\hat{H}(t_i)$ if $t_j>t_i$ and $\hat{H}(t_i)\hat{H}(t_j)$ otherwise. It gives rise to observable, nonlocal, time ordering effects if and only if $\left[\hat{H}(t_j),\hat{H}(t_i)\right ]\ne0$~\cite{kaplan,shakov}. Note that the non-trivial time dependence in $\hat U(t)$ arises from time dependent
$\hat H(t)$ and time ordering $\stackrel{\leftarrow}{T}$. 
\section{Measure of entanglement}
\label{concurrence}
For two-qubit systems, as an entanglement measure, Wootters concurrence is a well-defined quantity~\cite{wootters}. Its value ranges from 0 for a separable state to 1 for a maximally entangled~(Bell) state. The concurrence function is defined as:
\begin{equation}
\label{con1}
C(\hat{\rho})=\max\{0,\sqrt{\lambda_1}-\sqrt{\lambda_2}-\sqrt{\lambda_3}-\sqrt{\lambda_4}\},
\end{equation}
where $\lambda_i~(i=1,2,3,4)$ are the eigenvalues of the matrix $\hat{\rho}(t)(\hat{\sigma}_1^y\otimes\hat{\sigma}_2^y)\hat{\rho}(t)^*(\hat{\sigma}_1^y\otimes\hat{\sigma}_2^y)$ in descending order. Here $\hat{\rho}(t)$ is the density matrix of the qubits and $\hat{\rho}(t)^*$ is its complex conjugate. According to Schr\"{o}dinger equation~(\ref{hamiltonian}), it should be noted that the probability amplitudes $a_i(t)~(i=1,2,3,4)$ evolve in time interdependently. From this point, it is convenient to consider a general time-dependent two-qubit pure state in the form: $\left|\Psi(t)\right\rangle=a_1(t)\left|11\right\rangle+a_2(t)\left|10\right\rangle+a_3(t)\left|01\right\rangle+a_4(t)\left|00\right\rangle$. Then, it is straightforward to show that the concurrence function~(\ref{con1}) for the general pure state with density matrix $\hat{\rho}(t)=\left|\Psi(t)\right\rangle\left\langle\Psi(t)\right|$ is given by the simple equation as:
\begin{equation}
\label{conn}
C(\hat{\rho})=2\left|a_1(t)a_4(t)-a_2(t)a_3(t)\right|,
\end{equation}
where the time-dependent coefficients may be given by Eq.~(\ref{amps}) as:
\begin{equation}
\label{probability}
a_i(t)=\displaystyle\sum_{j=1}^4 U_{ij}(t) a_j(0),
\end{equation}
where $U_{ij}(t)~(i,j=1,2,3,4)$ are the matrix elements of $\hat{U}(t)$. According to Eqs.~(\ref{conn}) and~(\ref{probability}), to study entanglement dynamics of two qubits, one has to present the initial preparation of the qubits, i.e., $a_i(0)$, and the matrix elements of the time evolution matrix $\hat{U}(t)$. In this work, we will consider pure states as initial states. An initially pure state remains pure at all times under the dynamics given by Eq.~(\ref{hamiltonian}). It is well known that for pure states entanglement can quantify all quantum correlations between two two-level systems, while  such an identification is complicated for mixed states and entanglement signifies only a particular type of quantum correlation~\cite{lhvv}.
\section{Entanglement dynamics of two coupled qubits under the influence of three positive kicks}
\label{seckicked}
Depending on the complexity of the time evolution in  $\hat{H}_{int}(t)$, the dynamics may or may not be solved analytically. Most of the studies employ numerical solutions of the time evolution in order to control entanglement between two level quantum systems~\cite{ms,sla,wbsb,model2,wang,zhsk,zhsk2,fare}. However, analytic solutions are more convenient and easy to analyze, if they are available. For one qubit case, Shakov {\it et al.} listed some progressive approximations in which the time evolution may be solved analytically~\cite{shakov}. These approximations are the qubit having degenerate basis states, the external field being constant or changed slowly, the field being oscillating with a frequency close to the resonance frequency of the energy splitting of a qubit so that RWA solutions exist, the field being perturbative and a sudden or series of fast pulses~(kicks). The fast pulses are the most important limiting cases compared to the others because it was shown that they provide a full population transfer from one state to the other in a qubit~\cite{kaplan,shakov}. Moreover, the fast pulse and pulses have many potential implementations for some quantum tasks, such as NMR, quantum gates, excitation of electronic states in molecules, chemical reactions and quantum computing~\cite{jhm,vsbysc,slichter,krgt,swr,pk}.

We have obtained analytical expressions for the time evolution operator for one, two and three kick sequences and display the elements of $\hat{U}(t)$ in Appendices~A,~B and~C, respectively. They will be used in Eqs.~(\ref{conn}) and~(\ref{probability}) to calculate the dynamics of concurrence for various initial states. In Ref.~\cite{fare}, it was found that the entanglement of the qubits initially in $\left|01\right\rangle$ (or $\left|10\right\rangle$) and $\frac{1}{\sqrt{2}}(\left|10\right\rangle\pm\left|01\right\rangle)$ can be manipulated easily with an external field in $z$-direction, while for the initial states of the type $\left|11\right\rangle$, $\left|00\right\rangle$ or linear combination of them were immune to such a manipulation. So, we consider those states that cannot be affected by longitudinal external fields and analyze their dynamics under a transverse field.

In the following, we will discuss the effect of transverse field on the dynamics of the system which is initially in a separable, i.e., $\left|\Psi(0)\right\rangle=\left|11\right\rangle$ or in a maximally entangled state~i.e., $\left|\Psi(0)\right\rangle=\frac{1}{\sqrt{2}}(\left|11\right\rangle+\left|00\right\rangle)$. To see the effect of kicks on the entanglement dynamics for these initial states, one should note that before the kick the qubits evolve in accordance with the time independent Hamiltonian $\hat{H}_0$ and the propagator is given by 
\begin{eqnarray}
\label{freeevolution}
\hat{U}(t)&=&e^{-i\hat{H}_0 t}\nonumber\\
&=&\left[\begin{array}{cccc} e^{-i J t}  & 0 & 0  & 0 \\ 0  & e^{i J t} \cos(2 J t) &-i e^{i J t} \sin(2 J t)  & 0 \\ 0  & -i e^{i J t} \sin(2 J t) & e^{i J t} \cos(2 J t) & 0  \\ 0  & 0 & 0  & e^{-i J t}\end{array}\right].
\end{eqnarray}
Based on Eqs.~(\ref{conn}),~(\ref{probability}) and~(\ref{freeevolution}), the initially separable state remains separable, while the concurrence for the initial Bell state is equal to 1 at any time before the kick. That is, qubit-qubit interaction has no effect in the absence of external fields.
\begin{figure}[!hbt]\centering
{\scalebox{0.55}{\includegraphics{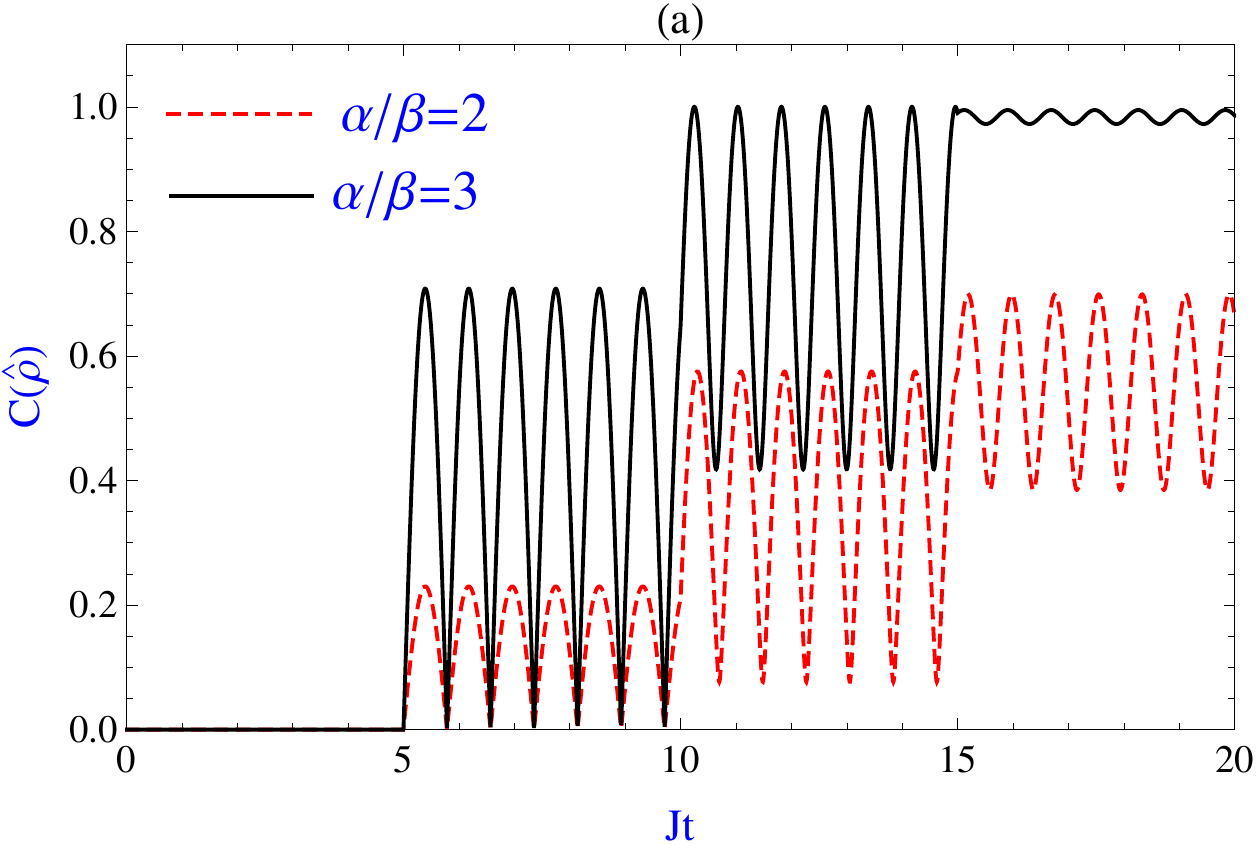}}}
{\scalebox{0.55}{\includegraphics{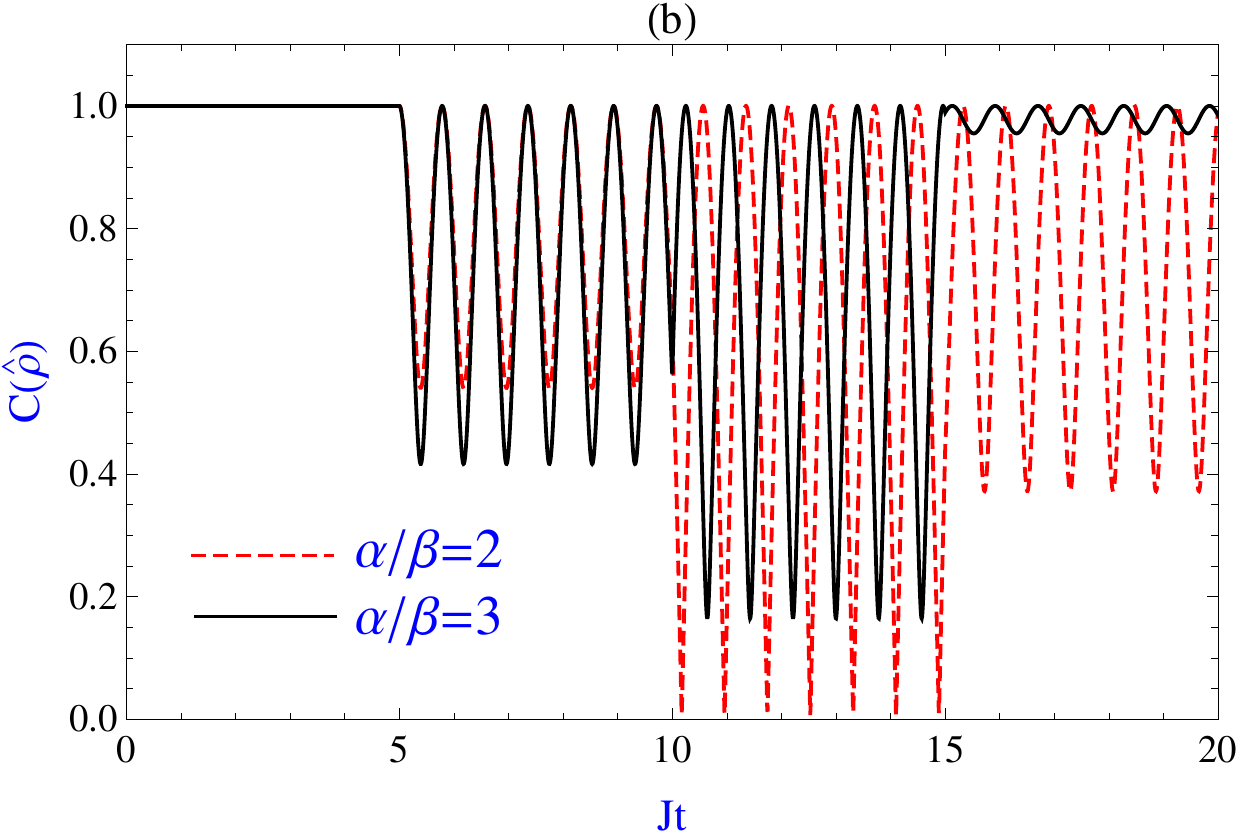}}}
\caption{$C(\hat{\rho})$ versus  $Jt$ for a sequence of three positive kicks applied  at times $T_1=5, T_2=10$ and $T_3=15$ for the initially pure states $\left|11\right\rangle$~(a) and $\frac{1}{\sqrt{2}}(\left|11\right\rangle+\left|00\right\rangle)$~(b) with $\theta=\frac{\pi}{2}$. Here the dashed lines correspond to $\alpha/\beta=2$ and the solid lines to $\alpha/\beta=3$, and we set $\beta=1$ and $J=1$.}
\end{figure}

Here we consider two qubits whose states are strongly perturbed by three positive kicks applied at times $t=T_1$, $t=T_2$ and $t=T_3$. The magnetic fields on qubits can be given as $B_1(t)=\alpha\displaystyle\sum_{i=1}^3\delta(t-T_i)$ and $B_2(t)=\beta\displaystyle\sum_{i=1}^3\delta(t-T_i)$, where $\alpha$ and $\beta$ are called integrated magnetic strengths~\cite{fare}, and $\delta(t)$ is the dirac delta function. For such a kick sequence, the integration over the time is trivial and the time evolution matrix~(\ref{U}) in the presence of time ordering can be obtained easily~\cite{fare,kaplan,shakov}. In the Appendix part, the evolution matrices after each kick are presented and by using these propagators in Figs.~1(a) and~1(b), the effects of the sequence of three kicks on $C(\hat{\rho})$ for $\left|11\right\rangle$ and $\frac{1}{\sqrt{2}}(\left|11\right\rangle+\left|00\right\rangle)$ initial states are displayed, respectively, with $\theta=\pi/2$ and $\alpha/\beta=2$ or $\alpha/\beta=3$. The concurrence for the considered initial states in the time domains $5<t<10$, $10<t<15$ and $t>15$ may be found by using Eqs.~(\ref{U^Kpara}),~(\ref{twokickpara}) and~(\ref{tkpara}), respectively, in Eqs.~(\ref{conn}) and~(\ref{probability}). Both figures show the pronounced effect of the kick on entanglement dynamics; the concurrence of these initial states starts oscillating just after a single kick with an increase in their oscillation amplitudes with the increase in the ratio $\alpha/\beta$~(see the time domain $5<t<10$). In fact, a sudden kick applied at $t=5$ induces entanglement from the initially unentangled qubits~(Fig.~1(a)) and yields oscillations in the concurrence function for the Bell state~(Fig.~1(b)). It is worth  mentioning here that for the magnetic fields acting in the $z$-axes, it was found that the qubits remain unentangled or maximally entangled state at all times for $\left|11\right\rangle$ or $\frac{1}{\sqrt{2}}(\left|11\right\rangle+\left|00\right\rangle)$ initial states despite the existence of strong external kicks~\cite{fare}. Comparing one, two and three kick regions, the sole effect of the number of kicks is found to change the amplitude~(or minimum) of $C(\hat{\rho})$ for initial Bell state. As can be seen from Fig.~1(b), after second kick, the amplitude of concurrence oscillations increases compared to that of first  kick, while it decreases after third kick. On the other hand, for  $\left|11\right\rangle$ state case, the effect of the number of kicks is complicated compared to the Bell state case, because each kick can change the amplitude, maximum and minimum of $C(\hat{\rho})$. It is obvious that after each kick the minimum of  $C(\hat{\rho})$ increases for initially separable state. The most important observation from these figures is the possibility of obtaining almost steady high concurrence around 1 after third kick for $\alpha/\beta=3$ and both initial states. This shows that for certain system parameters~(here we set $\alpha=3$, $\beta=1$, $J=1$ and $T=5$), it is possible to create highly entangled qubits from an initially separable state by perturbing the qubits via instantaneous pulses. Also note that after second kick, the minimum of $C(\hat{\rho})$ for Bell state can go to 0 for $\alpha/\beta=2$. As mentioned above, the qubit-qubit interaction has no effect on the entanglement of the considered states before the kick, while their oscillation frequencies depend on the qubit-qubit interaction strength, $J$, after the kick as can be seen from Figs.~1(a) and~1(b) as well as from the matrix elements in the Appendix part. 
\begin{figure}[!hbt]\centering
{\scalebox{0.5}{\includegraphics{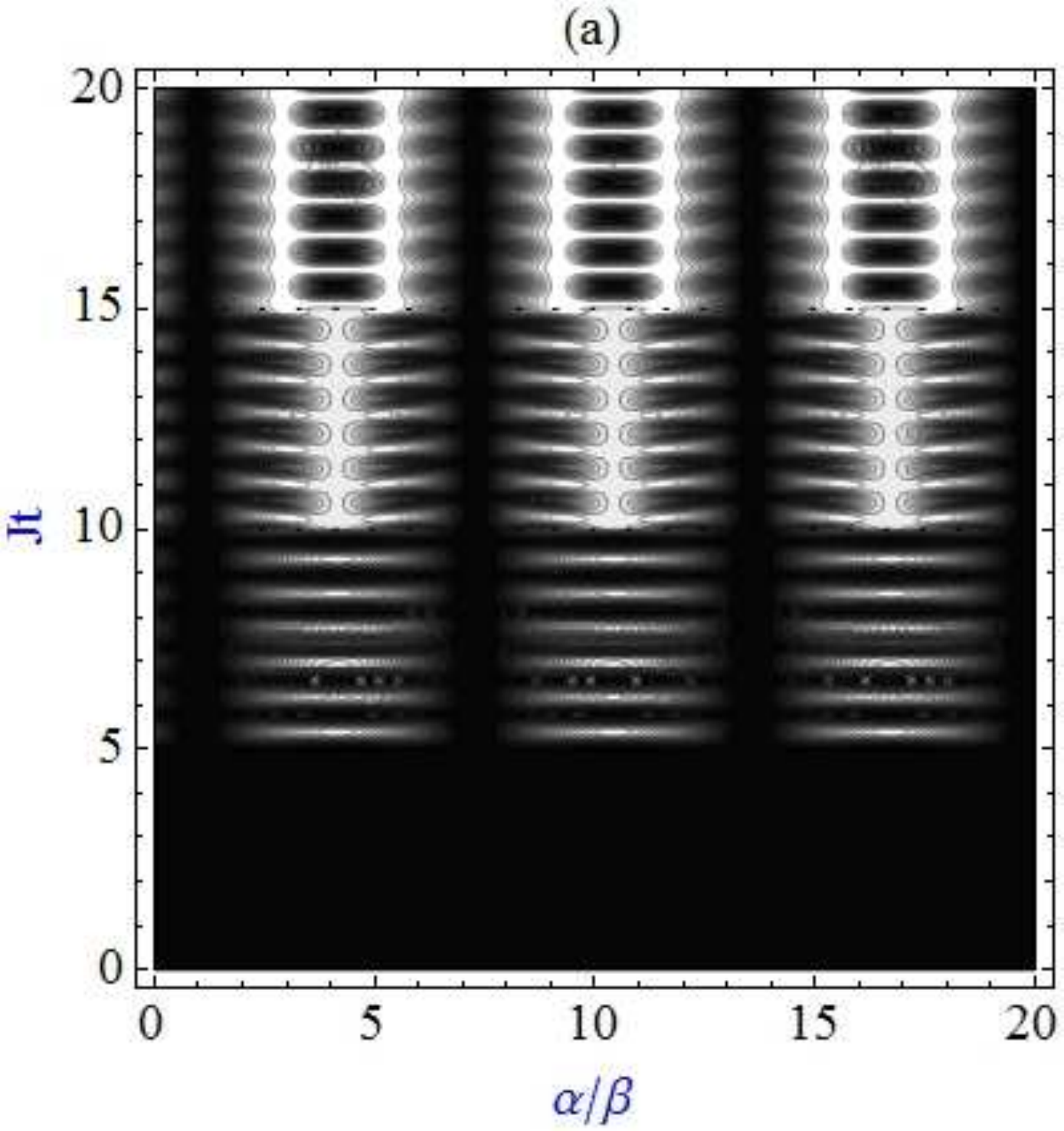}}}
{\scalebox{0.5}{\includegraphics{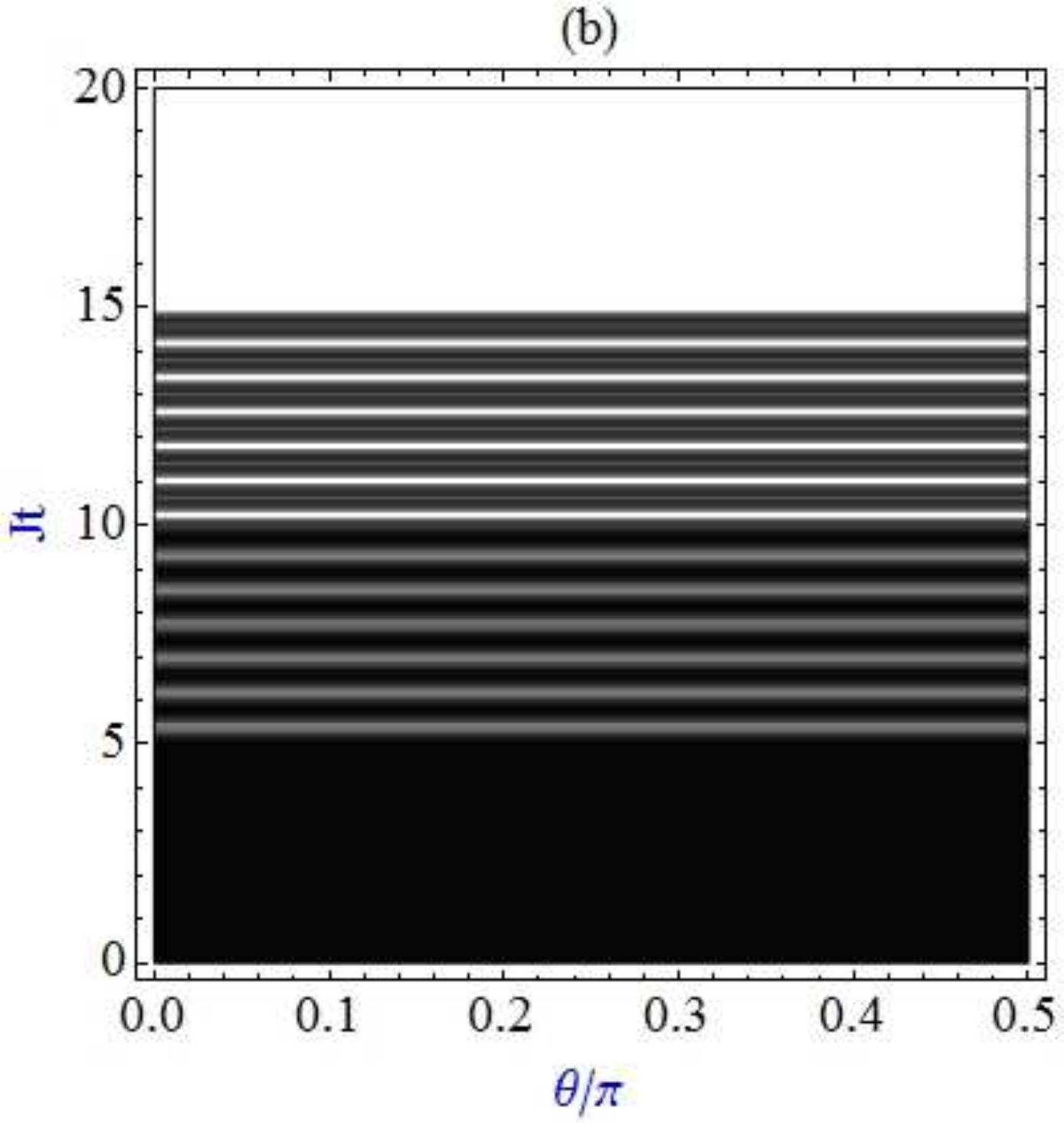}}}
\caption{(Colour online) (a) shows the contour plot of $C(\hat{\rho})$ versus $Jt$ and $\alpha/\beta$ with $\theta=\frac{\pi}{2}$. (b) shows the contour plot of $C(\hat{\rho})$  versus $Jt$ and $\theta/\pi$  with $\alpha/\beta=3$. Here the contour plots are for $\left|11\right\rangle$ initial state and include three positive kicks applied at times $T_1=5,T_2=10$ and $T_3=15$, and we set $\beta=1$ and $J=1$. In these contour plots we have assumed  twenty equidistant contours of concurrence between 0~(black) and 1~(white).}
\end{figure}

To further elucidate the effect of the strength and the direction of the magnetic fields on $C(\hat{\rho})$, we have displayed the contour plot of $C(\hat{\rho})$ versus $Jt$ and $\alpha/\beta$ for the separable initial state in Fig.~2(a) and Bell state in Fig.~3(a) with $\theta=\pi/2$, while in Figs.~2(b) and~3(b), $C(\hat{\rho})$ versus $Jt$ and $\theta/\pi$ have been plotted for $\left|11\right\rangle$ and  $\frac{1}{\sqrt{2}}(\left|11\right\rangle+\left|00\right\rangle)$ with $\alpha/\beta=3$, respectively. In these figures, twenty equidistant contours of concurrence between 0~(black) and 1~(white) are shown. From Fig.~2(b), it can be deduced that the entanglement dynamics for the initially separable state is independent of the direction of external fields as long as the field is transverse, which also can be seen from the relevant time evolution matrix elements in Appendices where the concurrence is $\theta$-independent. Fig.~2(a) shows two pronounced results for $\alpha/\beta$-dependence of the concurrence for $\left|11\right\rangle$ state: i) The separable state is found to remain almost separable despite the strong external kicks for $\alpha/\beta=1$ and $\alpha/\beta\cong7.25,13.25,19.5$. ii) The long lived high entanglement regions which are indicated in white straight stripes sections have long lifetimes only after the second and the third kicks. From the comparison of  the one, two and three positive kick sections, it seems that each kick widens the long lived high entanglement $\alpha/\beta$-area. This shows the necessity of using multiple kicks to create almost steady high entanglement. Also note that each kick creates a different $\alpha/\beta$ oscillatory structure for $C(\hat{\rho})$.
\begin{figure}[!hbt]\centering
{\scalebox{0.5}{\includegraphics{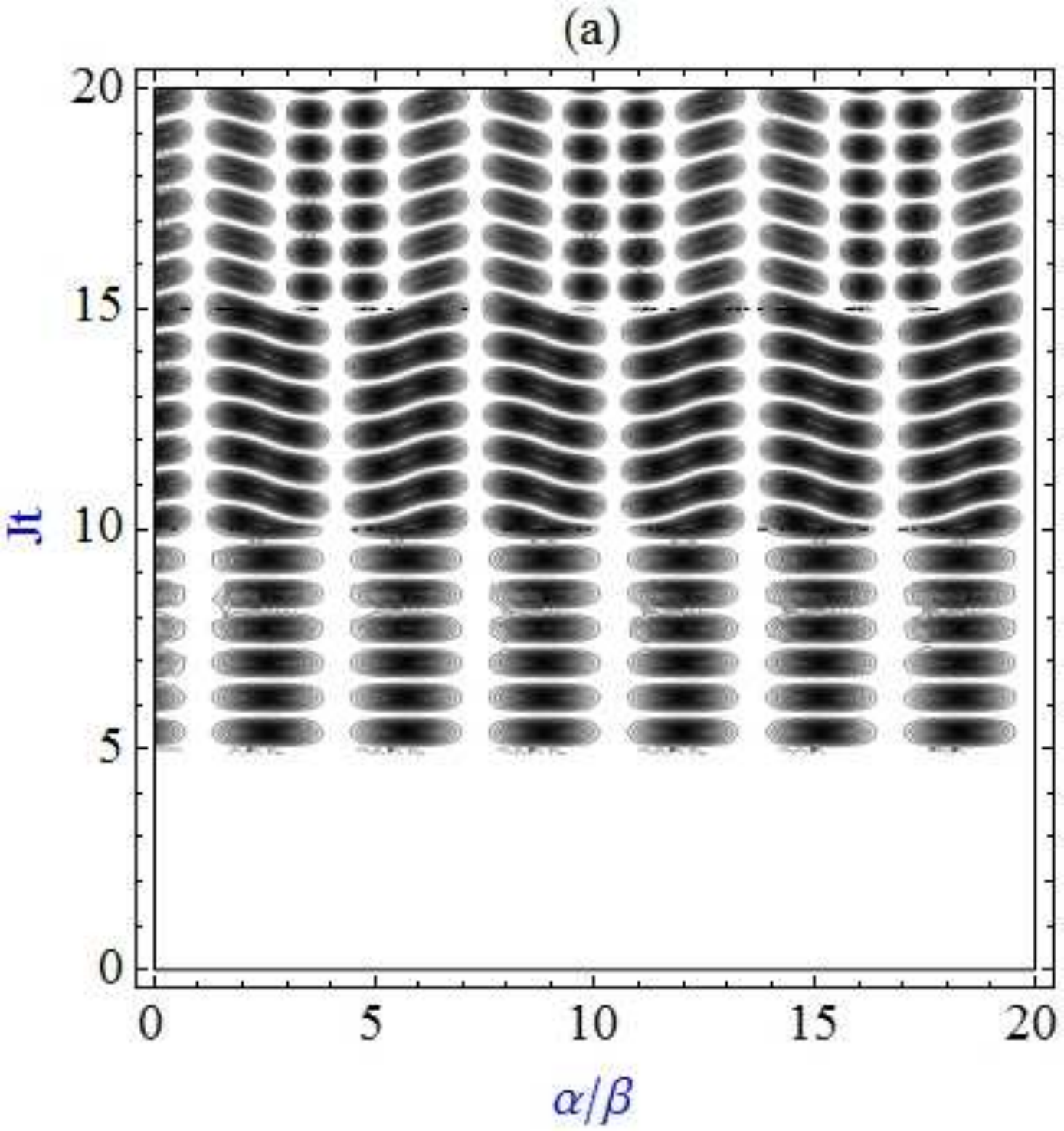}}}
{\scalebox{0.5}{\includegraphics{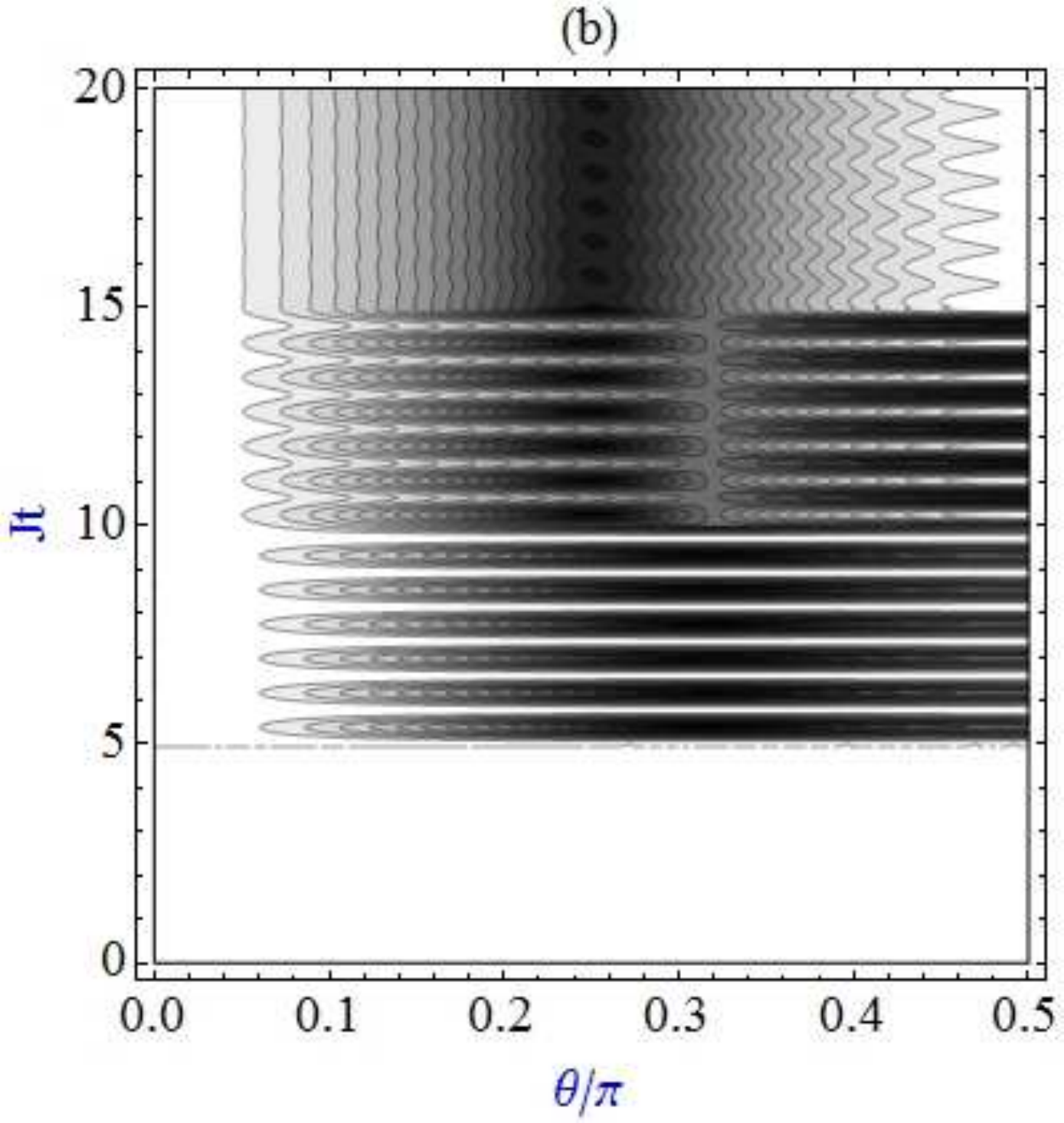}}}
\caption{(Colour online) (a) shows the contour plot of $C(\hat{\rho})$ versus $Jt$ and $\alpha/\beta$ with $\theta=\frac{\pi}{2}$. (b) shows the contour plot of $C(\hat{\rho})$  versus $Jt$ and $\theta/\pi$  with $\alpha/\beta=3$. Here the contour plots are for $\frac{1}{\sqrt{2}}(\left|11\right\rangle+\left|00\right\rangle)$ initial state and include three positive kicks applied at times $T_1=5,T_2=10$ and $T_3=15$, and we set $\beta=1$ and $J=1$. In these contour plots we have assumed  twenty equidistant contours of concurrence between 0~(black) and 1~(white).}
\end{figure}

Contrary to the initially separable state case, the concurrence for the initial Bell state strongly depends on the direction of the external magnetic fields if $\theta/\pi>0.04$ as can be seen from Fig.~3(b); for $\theta/\pi<0.04$, the entanglement between two qubits is almost unperturbed by the external positive kicks. Similar to $\left|11\right\rangle$ state case, the concurrence for $\frac{1}{\sqrt{2}}(\left|11\right\rangle+\left|00\right\rangle)$ is found to be unaffected from three kicks for $\alpha/\beta=1$ and $\alpha/\beta\cong4.25,7.25,10.5,13.5,16.75,19.75$ that can be deduced from Fig.~3(a). This figure also shows that every kick region has a different $\alpha/\beta$ periodic structure and long-lived high entanglement can be effectively obtained and controlled only after third positive kick for the initial Bell state after perturbing the entanglement dynamics of qubits with external kicks, for example in the time domain $15<t<20$ for $\alpha/\beta\cong2.75,5.25,9.25,11.75,15.5,18.0$. From Fig.~3(b), one can see that it is possible to get a constant long-lived entanglement with desired magnitude after third kick by adjusting $\theta$. For example, $C(\hat{\rho})\approx 0.93$, $C(\hat{\rho})\approx 0.44$, $C(\hat{\rho})\approx 0.08$ and $C(\hat{\rho})\approx 0.63$  for $\theta/\pi=0.06$, $\theta/\pi=0.18$, $\theta/\pi=0.25$ and $\theta/\pi=0.36$, respectively.
\section{Entanglement dynamics of two qubits under the influence of Gaussian pulses}
\label{gaussian}
The kicked approximation  is based on the energy level of the qubit, $\Delta E$, and the width of the pulse, $\tau$, and is valid if $\Delta E\tau<<1$. Depending on the physical implementation of the qubit, it might be difficult to obtain an external field as a delta function kick. Instead of kicks, Gaussian pulses can be used. As is well known, the kicked approximation is the limiting case of Gaussian pulses~$\left(i.e.,  \lim_{\tau\rightarrow0}\frac{\alpha_K}{\sqrt{\pi}\tau}e^{-\frac{(t-T_K)^2}{\tau^2}}=\alpha_K\delta(t-T_K)\right)$, thus Gaussian pulses enable us to consider the effects of using finite-width pulses on entanglement dynamics. From this point, in this section, we present the results of numerical calculations of concurrence for two qubits whose states are strongly perturbed by series  of three positive narrow Gaussian pulses for the initially separable $\left|11\right\rangle$ and maximally entangled $\frac{1}{\sqrt{2}}(\left|11 \right\rangle+\left|00\right\rangle)$ states. The concurrence for the considered initial states may be calculated by solving numerically the set of first order coupled ordinary differential equations in Eq.~(\ref{hamiltonian}):
\begin{eqnarray}
\label{num}
i\dot{a}_1(t)&=&Ja_1(t)+\frac{1}{2}e^{-i\theta}B_2(t)a_2(t)+\frac{1}{2}e^{-i\theta}B_1(t)a_3(t),\nonumber\\ 
i\dot{a}_2(t)&=&\frac{1}{2}e^{i\theta}B_2(t)a_1(t)-J a_2(t)+2J a_3(t)+\frac{1}{2}e^{-i\theta}B_1(t)a_4(t),\nonumber\\
i\dot{a}_3(t)&=&\frac{1}{2}e^{i\theta}B_1(t)a_1(t)+2J a_2(t)-J a_3(t)+\frac{1}{2}e^{-i\theta}B_2(t)a_4(t),\nonumber\\
i\dot{a}_4(t)&=&\frac{1}{2}e^{i\theta}B_1(t)a_2(t)+\frac{1}{2}e^{i\theta}B_2(t)a_3(t)+J a_4(t),
\end{eqnarray}
with replacing $B_1(t)\rightarrow\frac{\alpha}{\sqrt{\pi}\tau}\displaystyle\sum_{i=1}^3e^{-\frac{(t-T_i)^2}{\tau^2}}$ and $B_2(t)\rightarrow\frac{\beta}{\sqrt{\pi}\tau}\displaystyle\sum_{i=1}^3e^{-\frac{(t-T_i)^2}{\tau^2}}$. Here the Gaussian pulses are assumed to have the same width $\tau$ and centered at times $t=T_1,~T_2$ and $T_3$.  
We will determine how the concurrences of the initially separable and Bell states depend on the pulse width. It should be noted here that if $\tau$ were chosen small enough, the results obtained in the previous section could be reached. 
\begin{figure}[!hbt]\centering
{\scalebox{0.45}{\includegraphics{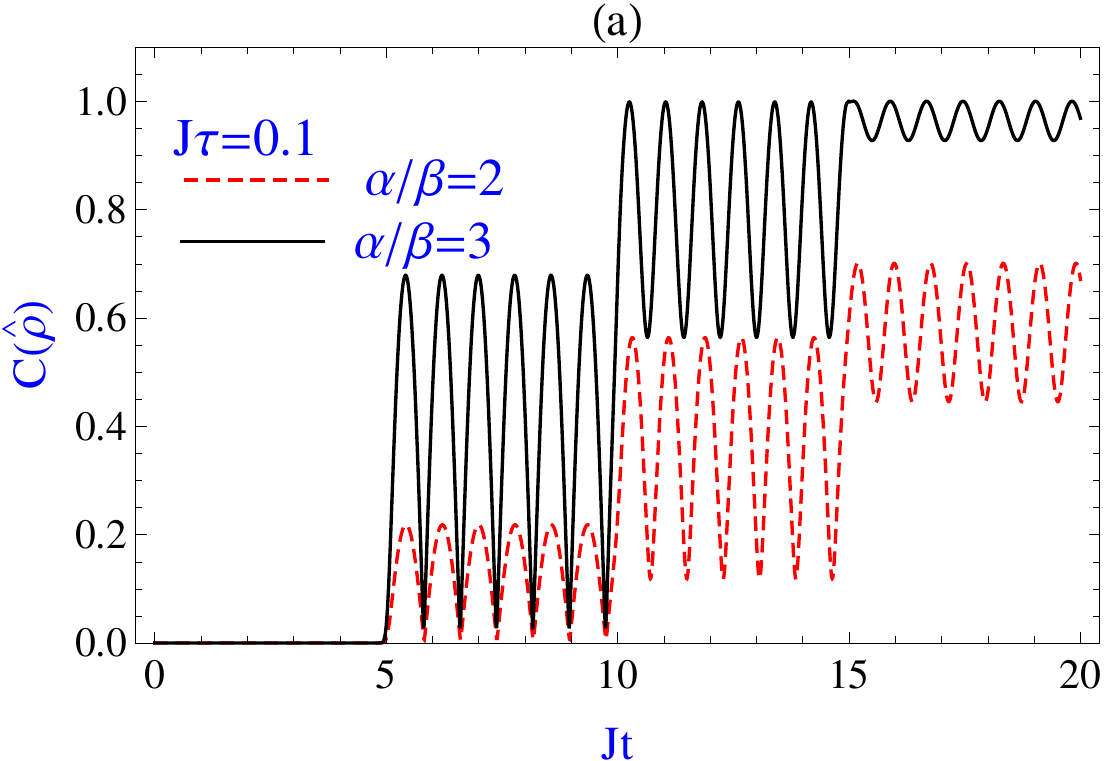}}}
{\scalebox{0.45}{\includegraphics{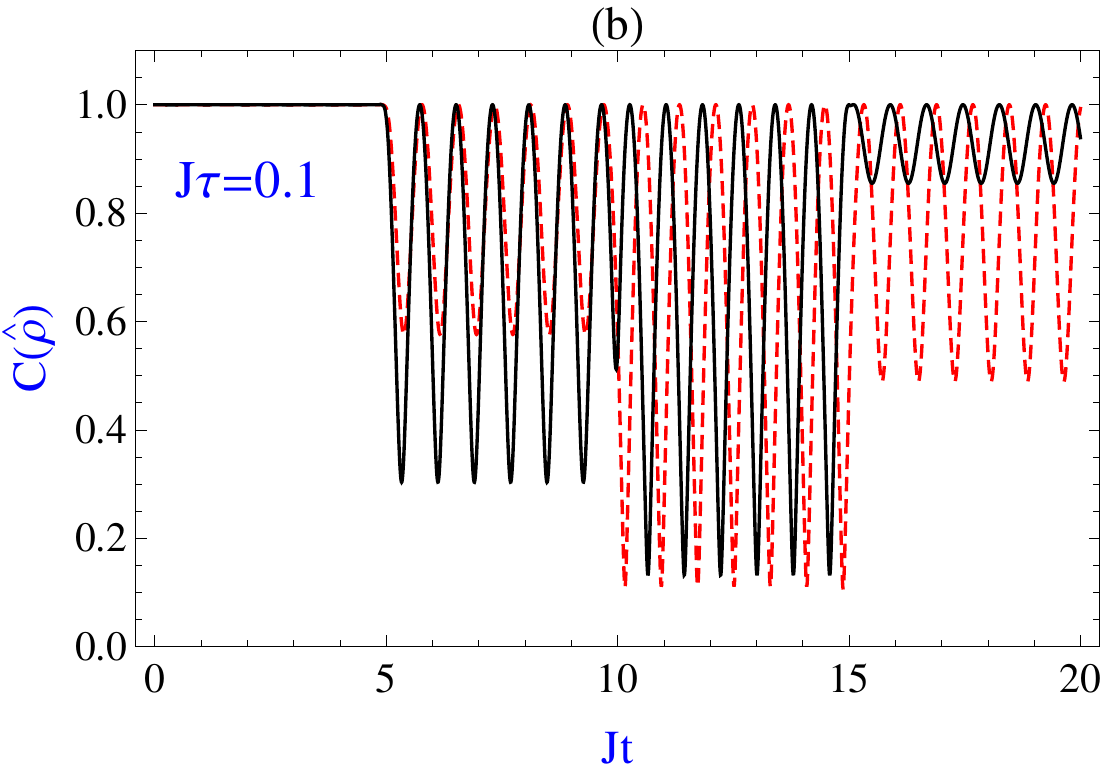}}}

{\scalebox{0.45}{\includegraphics{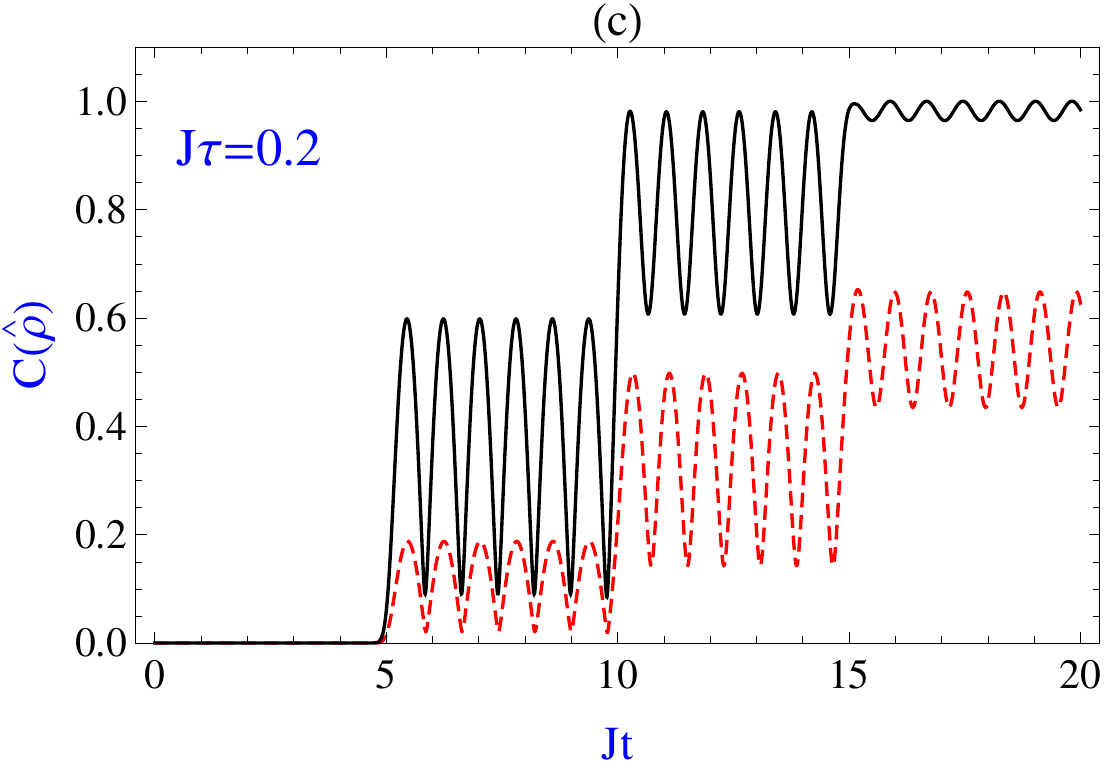}}}
{\scalebox{0.45}{\includegraphics{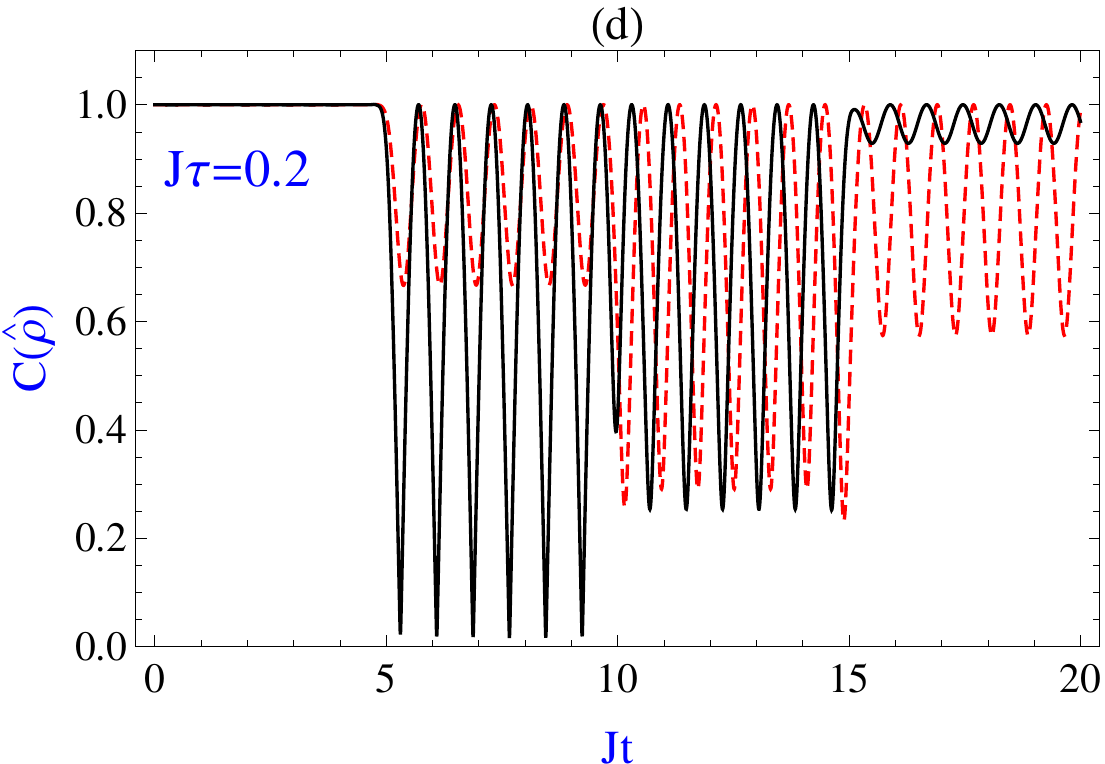}}}

{\scalebox{0.45}{\includegraphics{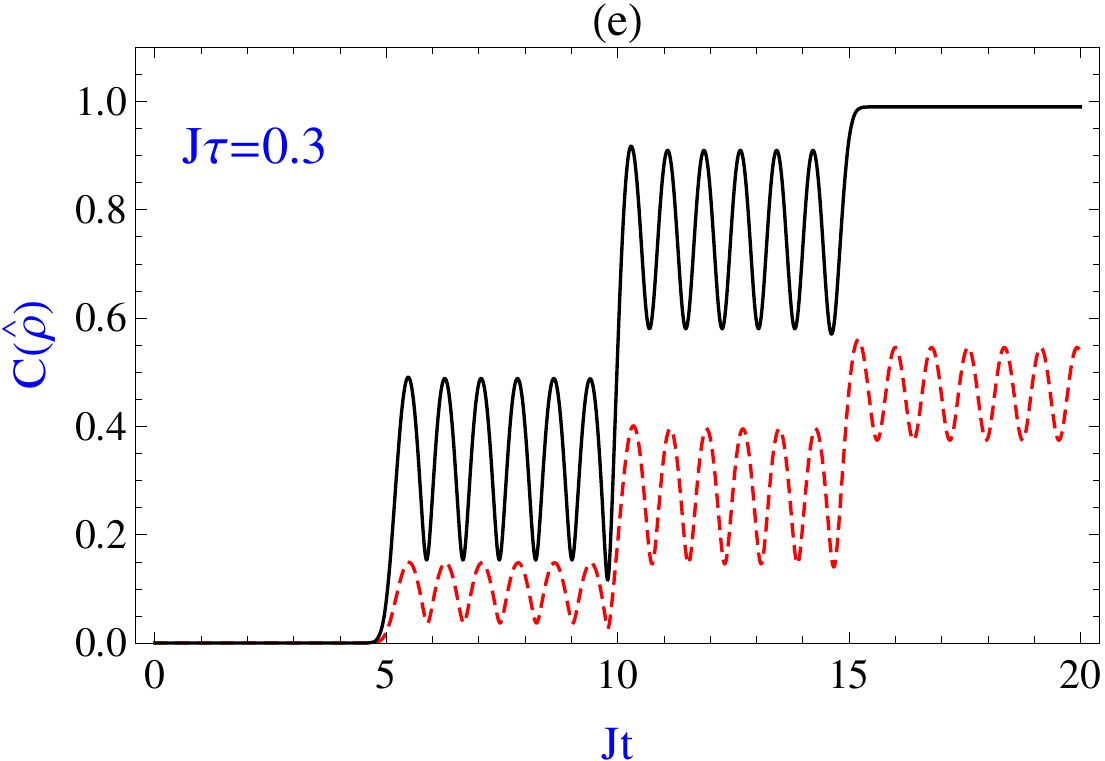}}}
{\scalebox{0.45}{\includegraphics{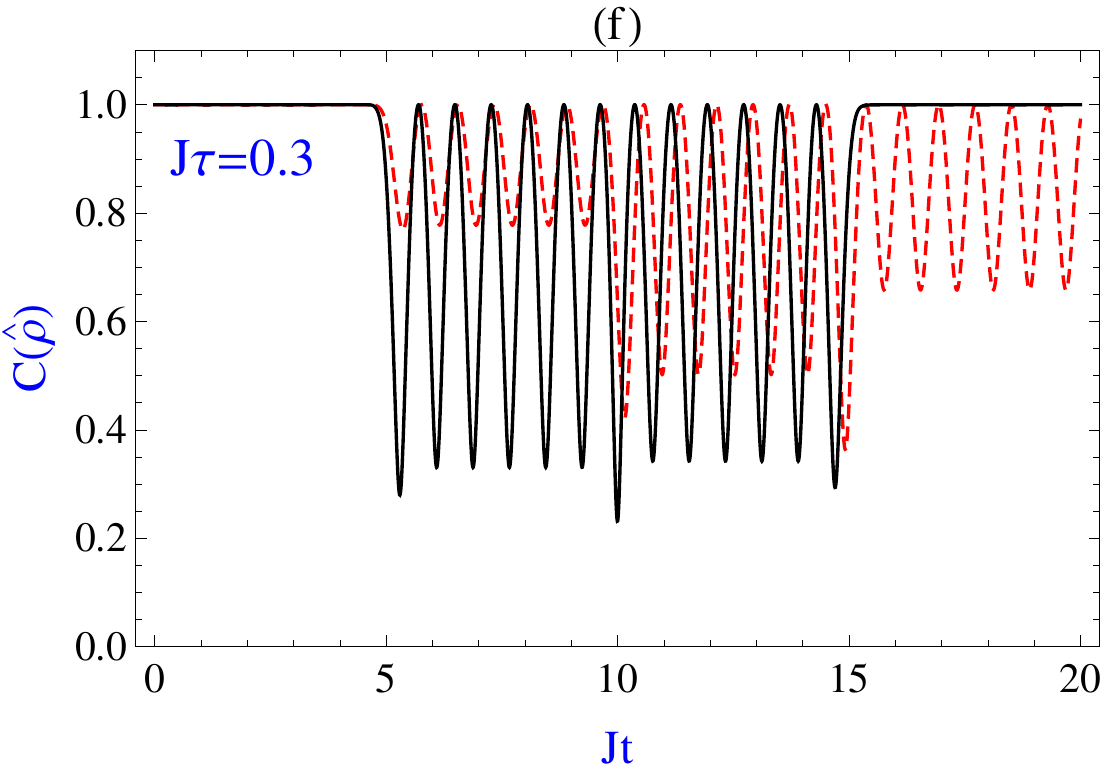}}}
\caption{$C(\hat{\rho})$ versus  $Jt$ for a sequence of three positive Gaussian pulses centered at $T_1=5, T_2=10$ and $T_3=15$ for the initially pure states $\left|11\right\rangle$~((a),~(c) and~(e)) and $\frac{1}{\sqrt{2}}(\left|11 \right\rangle+\left|00\right\rangle)$~((b),~(d) and~(f)) with  $\theta=\frac{\pi}{2}$. Here the dashed lines correspond to $\alpha/\beta=2$ and the solid lines to $\alpha/\beta=3$, and we consider three dimensionless pulse widths as $J\tau=0.1$~((a) and (b)), $J\tau=0.2$~((c) and (d)) and $J\tau=0.3$~((e) and (f)), and we set $J=1$ and $\beta=1$.}
\end{figure}

In Figs.~4(a)-4(f), we have shown  $C(\hat{\rho})$ versus  $Jt$ for a system strongly perturbed by three narrow Gaussian pulses  for $\left|11\right\rangle$ and $\frac{1}{\sqrt{2}}(\left|11\right\rangle+\left|00\right\rangle)$ initial states with $\theta=\frac{\pi}{2}$ and $\alpha/\beta=2$ or $\alpha/\beta=3$. Comparing Figs.~1(a) and~1(b) with the subfigures in Fig.~4, the width of the pulse changes the minimum, maximum and the amplitude of $C(\hat{\rho})$ for $\left|11\right\rangle$ state, while for Bell state, this effect is in its minimum~(or amplitude).  The most pronounced observation from these figures is the existence of constant high concurrence nearly 1 at times $t>15$ for $\alpha/\beta=3$ and $J\tau=0.3$ for both initial states as can be seen from the solid lines in Figs.~4(e) and~4(f). From Fig.~4(e), it is safe to deduce that with a sufficiently high pulse width and $\alpha/\beta$ ratio, it is possible to obtain a state very close to Bell state from an initially separable state by using a Gaussian pulse sequence.
\begin{figure}[!hbt]\centering
{\scalebox{0.5}{\includegraphics{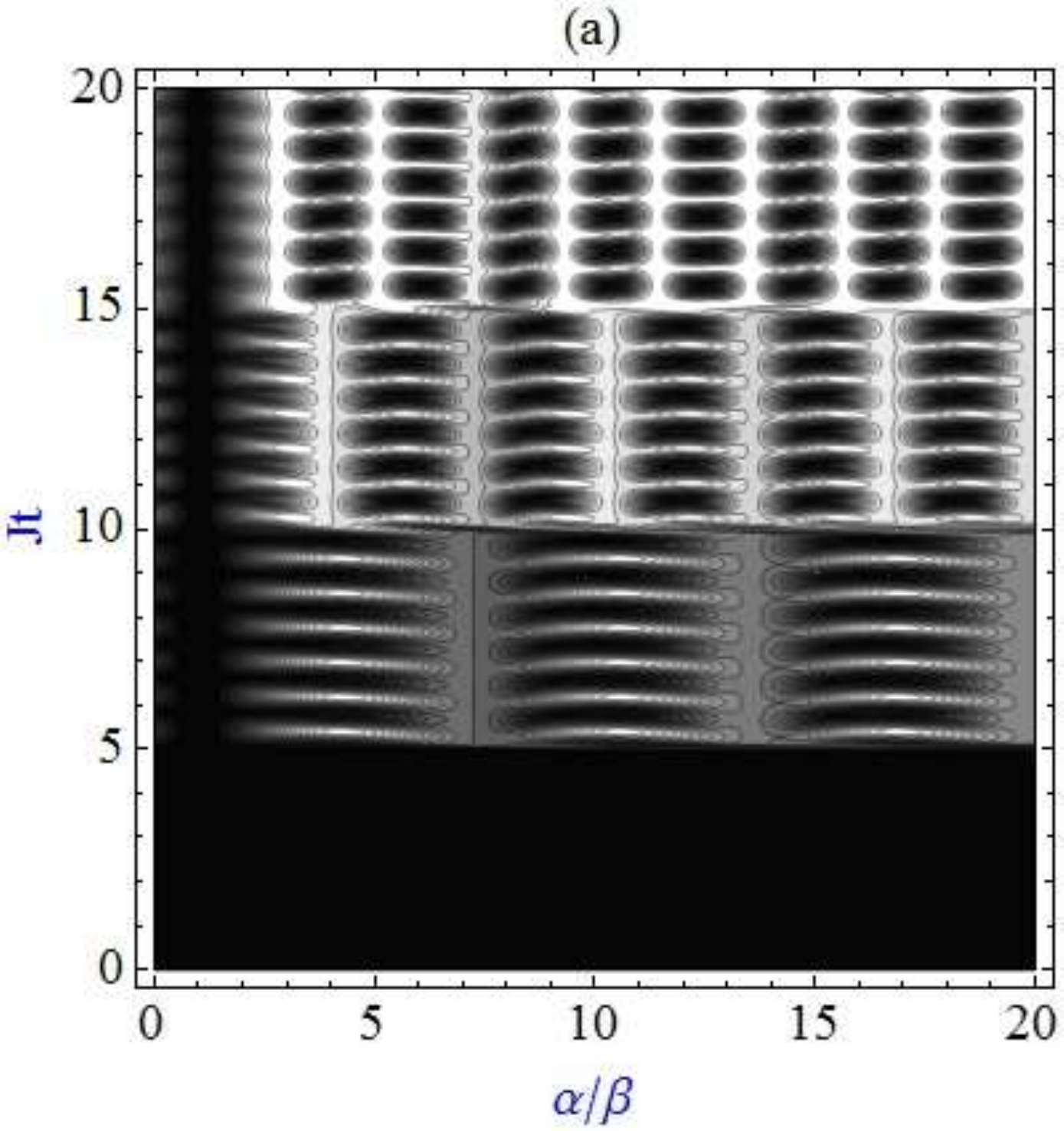}}}
{\scalebox{0.5}{\includegraphics{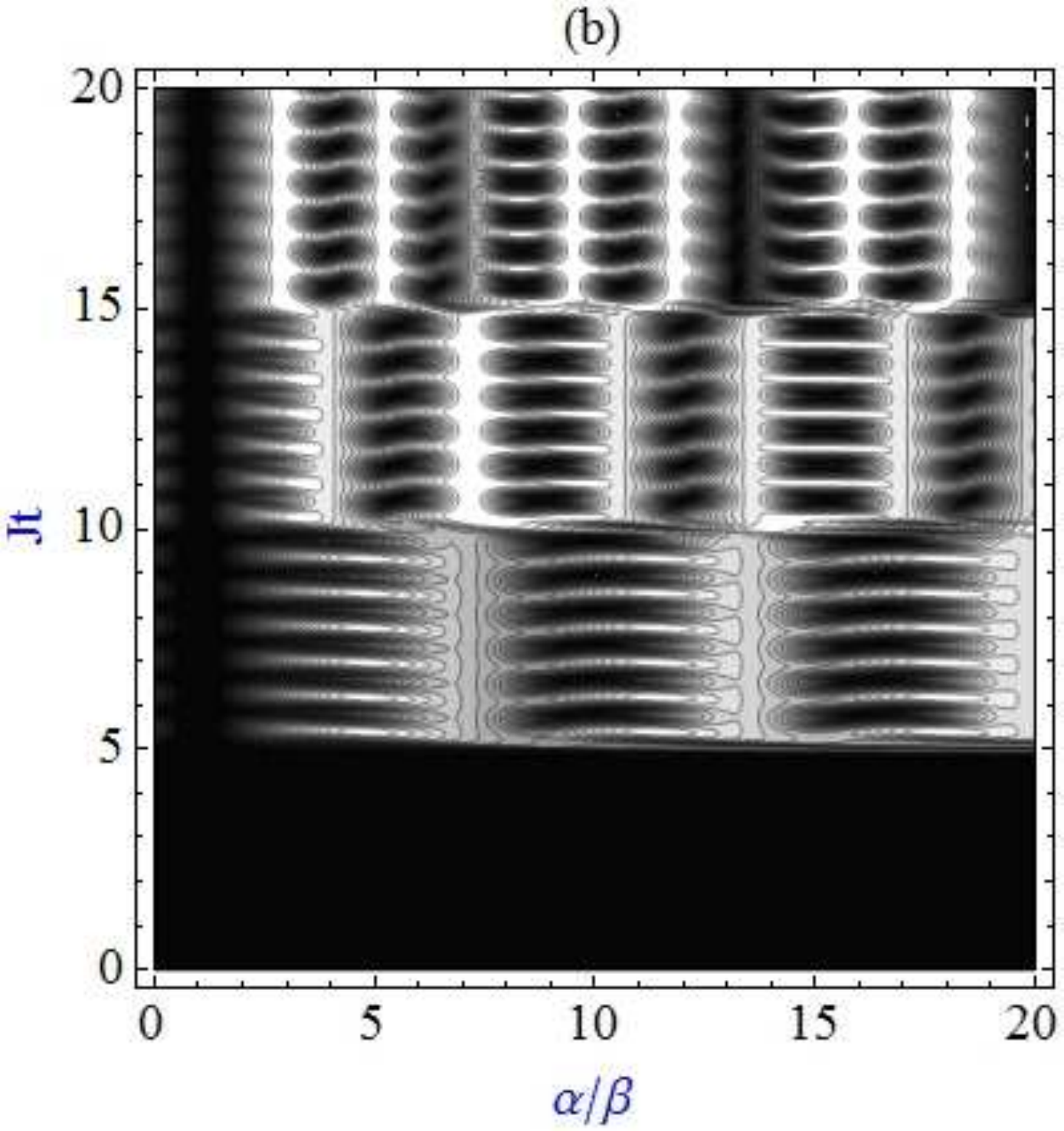}}}

{\scalebox{0.5}{\includegraphics{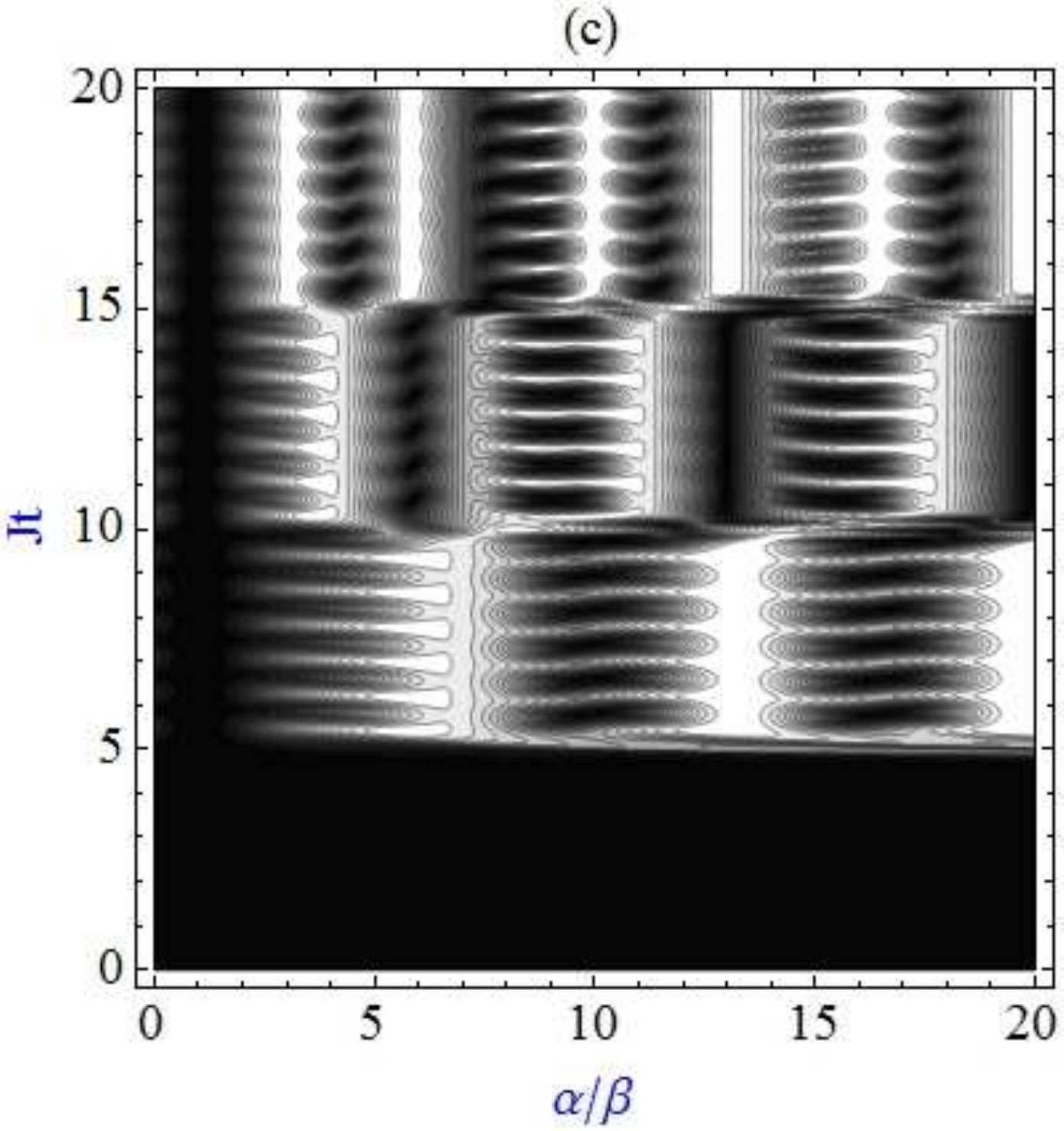}}}
\caption{(Colour online) The contour plot of $C(\hat{\rho})$ versus $Jt$ and $\alpha/\beta$ with $\theta=\frac{\pi}{2}$ and $J\tau=0.1$~(a), $J\tau=0.2$~(b) and $J\tau=0.3$~(c).  Here the contour plots are for $\left|11\right\rangle$ initial state and include three positive Gaussian pulses centered at times $T_1=5,T_2=10$ and $T_3=15$, and we set $J=1$ and $\beta=1$. In these contour plots we have assumed  twenty equidistant contours of concurrence between 0~(black) and 1~(white).}
\end{figure}

To further elucidate the interplay between  a pulse sequence, relative magnetic strength on qubits, pulse width and entanglement, we display the contour plot of  $C(\hat{\rho})$ as function of $Jt$ and $\alpha/\beta$ at three different pulse widths $J\tau=0.1, 0.2, 0.3$ in Figs.~5(a)-5(c). Comparing the subfigures in Fig.~5 with the ideal kick case~(Fig.~2(a)), the most important effect of the increasing pulse width is found to be enlargement of the area of the long lived high entanglement regions which are the white straight areas toward the top of Figs.~5(a)-5(c). For the kick case, it was concluded that the long lived high entanglement regions only appear after second and third positive kicks, while for  wider Gaussian pulses, these regions also appear after the first pulse~(for example, see the region $5<t<10$ in Fig.5(c)). Contrary to Fig.~2(a), $\alpha/\beta$ periodicity does not exist in Figs.~5(a)-5(c); increasing the pulse width coalesces the  $\alpha/\beta$-dependent oscillatory structure and produce non-periodic structures. Furthermore, $\left|11\right\rangle$ state remains separable if and only if $\alpha/\beta=1$ under the strong influence of Gaussian pulses. On the other hand, when $\tau\rightarrow0$~(see Fig.~2(a)), this holds for  $\alpha/\beta=1$ as well as for $\alpha/\beta\cong7.25,13.25,19.5$, as was the case for the kick sequence. Peculiarly, Fig.~5(c) shows that there is a sudden transition between long lasting maximally entangled and almost separable states of the system just after a pulse for $\alpha=13\beta$ and $\alpha=19.8\beta$. Moreover, the subfigures in Fig.~5 demonstrates the possibility of obtaining desired values of steady entanglement by carefully designed pulse or pulse sequence and system parameters.
\begin{figure}[!hbt]\centering
{\scalebox{0.5}{\includegraphics{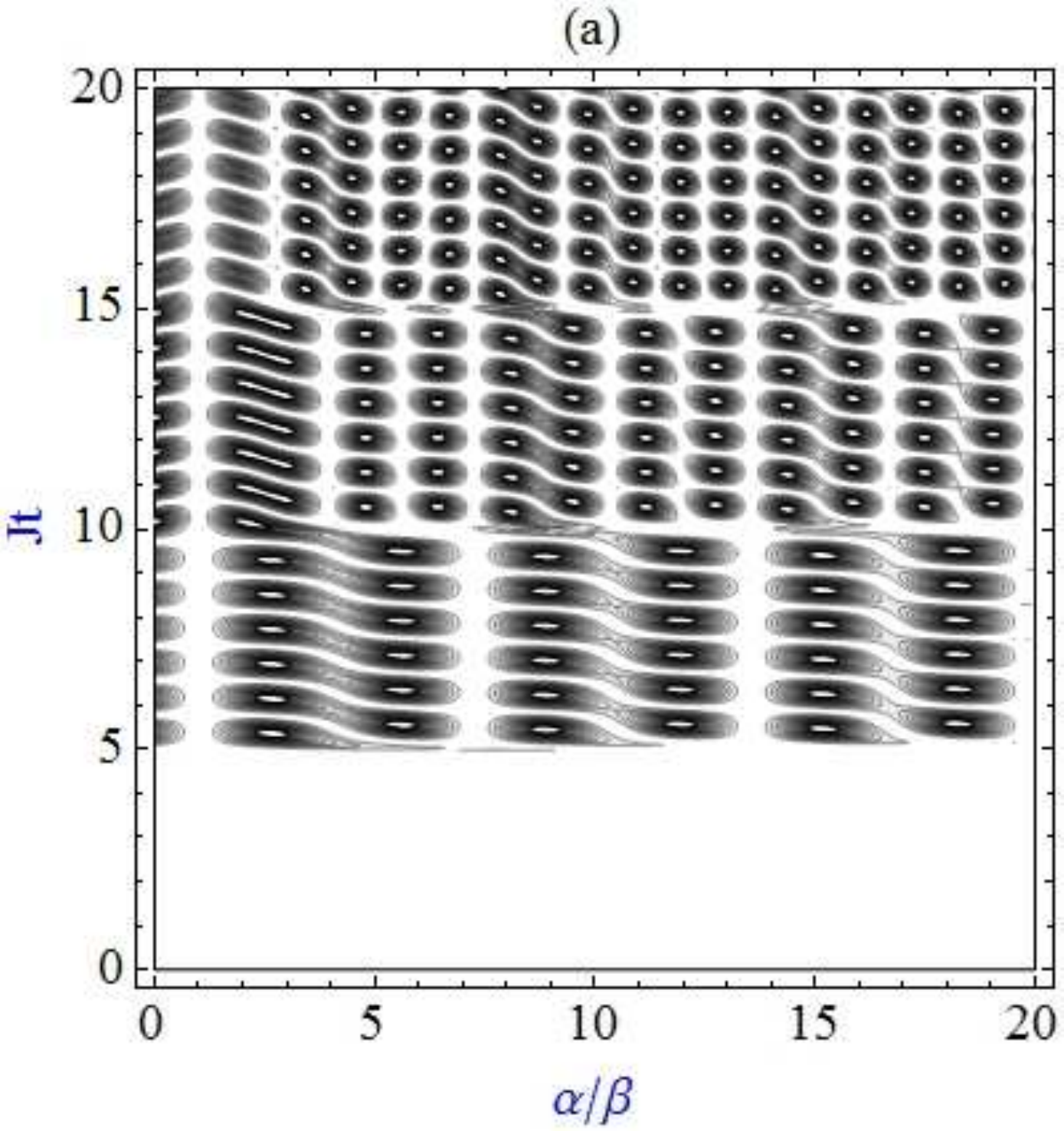}}}
{\scalebox{0.5}{\includegraphics{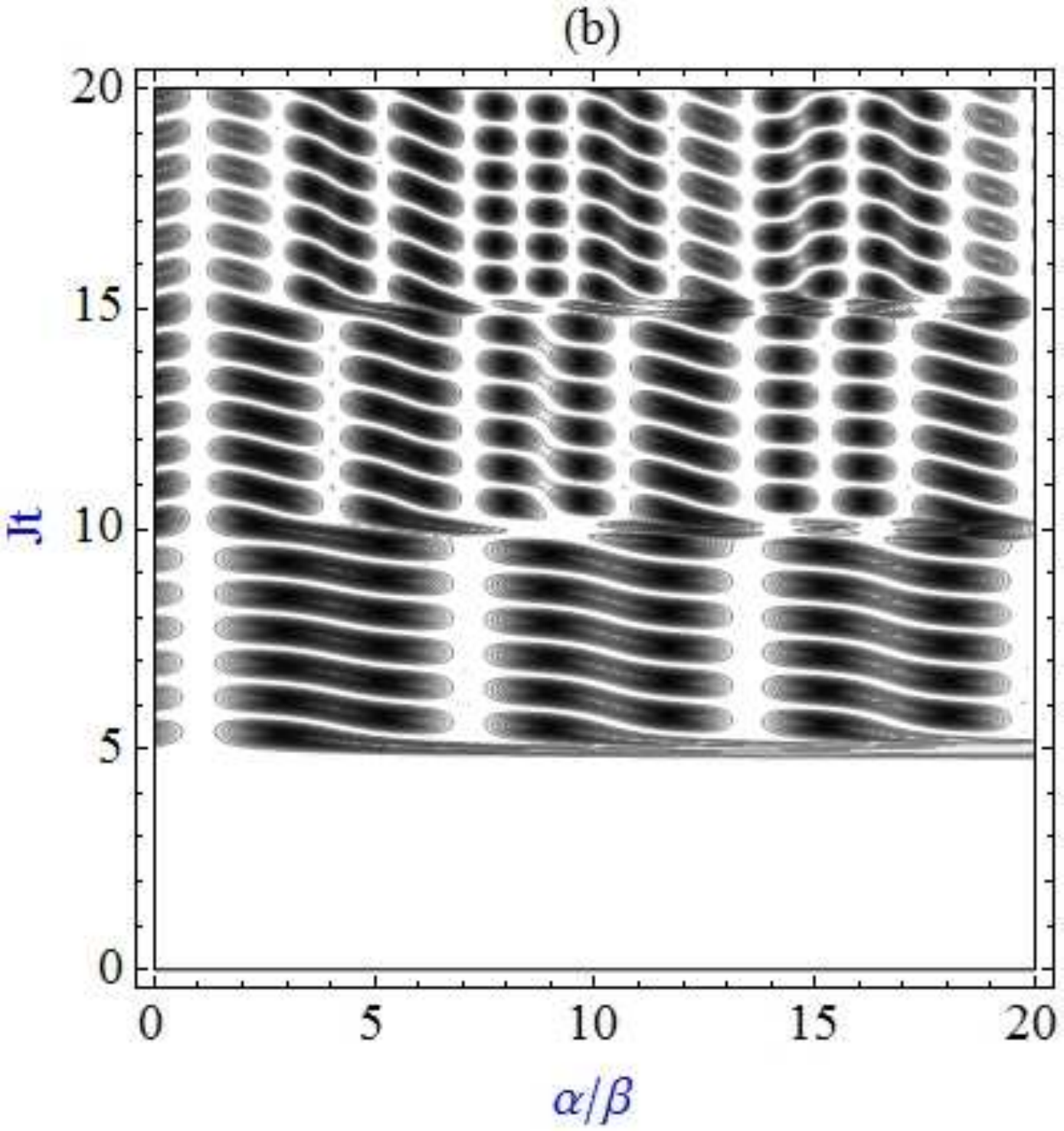}}}

{\scalebox{0.5}{\includegraphics{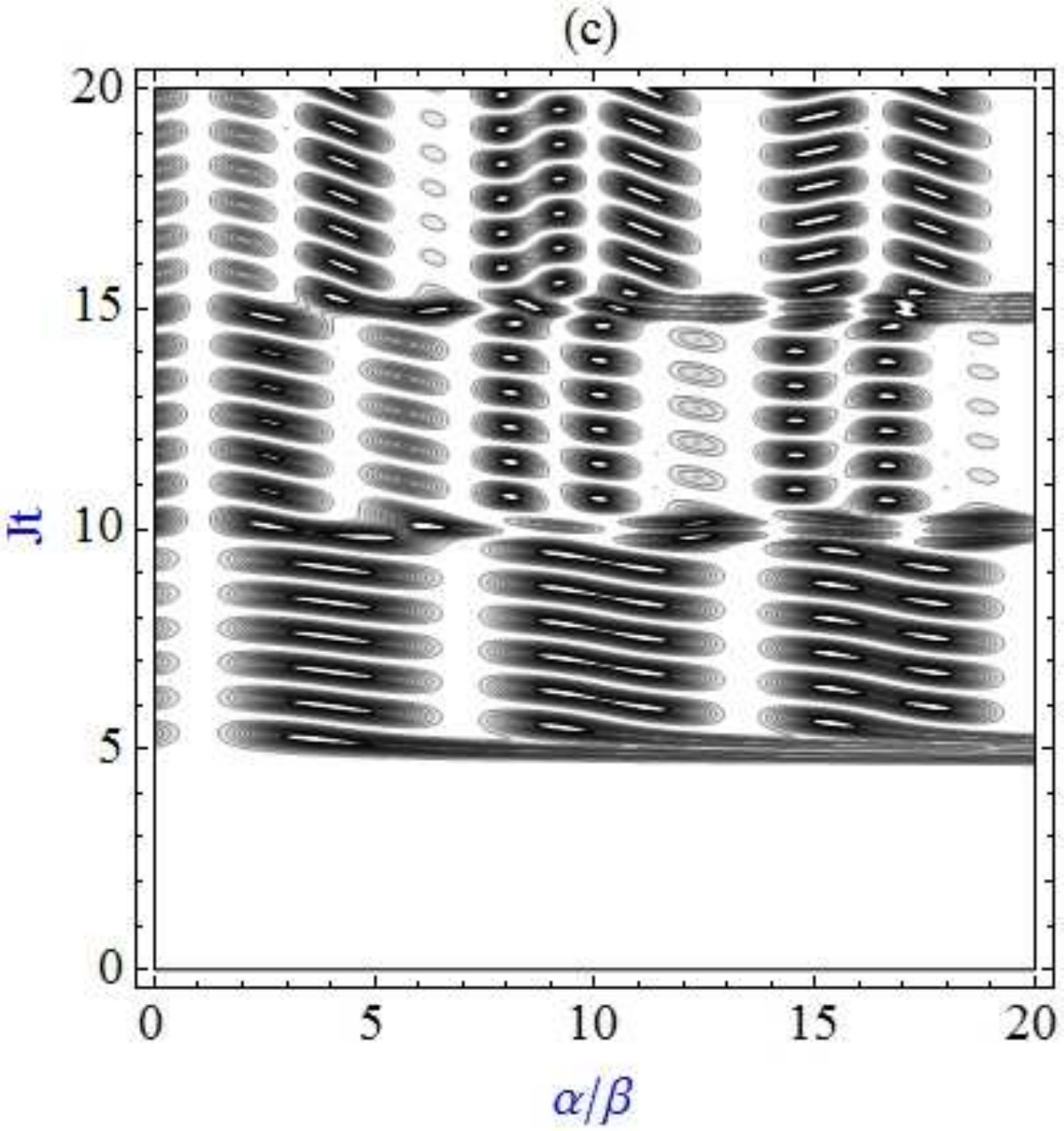}}}
\caption{(Colour online) The contour plot of $C(\hat{\rho})$ versus $Jt$ and $\alpha/\beta$ with $\theta=\frac{\pi}{2}$ and $J\tau=0.1$~(a), $J\tau=0.2$~(b) and $J\tau=0.3$~(c).  Here the contour plots are for $\frac{1}{\sqrt{2}}(\left|11\right\rangle+\left|00\right\rangle)$ initial state and include three positive Gaussian pulses centered at times $T_1=5,T_2=10$ and $T_3=15$, and we set $J=1$ and $\beta=1$. In these contour plots we have assumed  twenty equidistant contours of concurrence between 0~(black) and 1~(white).}
\end{figure}

In Fig.~6, the contour plot of $C(\hat{\rho})$ versus $Jt$ and $\alpha/\beta$ are displayed for $\frac{1}{\sqrt{2}}(\left|11\right\rangle+\left|00\right\rangle)$ initial state for the system under three Gaussian pulses having dimensionless pulse widths as $J\tau=0.1,0.2$ and 0.3. Similar to $\left|11\right\rangle$ state case, $C(\hat{\rho})$ of Bell state is undisturbed by the highly wider Gaussian pulses if and only if $\alpha/\beta=1$. In Fig.~3(a), we have noted that the long lived high entanglement regions can only be obtained after third kick after disturbing the entanglement dynamics for the Bell state, while these regions also appears after applying first and second Gaussian pulses as can be seen from Figs.~6(b) and~6(c). 

Note that we have not displayed the contour plots of $C(\hat{\rho})$ versus $\theta/\pi$ and $Jt$ for the Gaussian pulse magnetic fields, because they provide no extra information compared to the ideal kick case; for the separable $\left|11\right\rangle$ initial state,  $C(\hat{\rho})$ is independent of $\theta$ regardless of the value of the pulse width $\tau$. For the initial Bell state, the overall $\theta$-dependence is very similar to that of the kick sequence control.

We have also considered $\left|00\right\rangle$ separable, $\left|\Phi_-\right\rangle=\frac{1}{\sqrt{2}}(\left|11\right\rangle-\left|00\right\rangle)$ and $\left|\Psi_\pm\right\rangle=\frac{1}{\sqrt{2}}(\left|10\right\rangle\pm\left|01\right\rangle)$ Bell states as initial states and obtained some interesting dynamics. The dynamics of  $\left|00\right\rangle$ initial state under control sequences considered in the present study are found to be exactly same as that of  $\left|11\right\rangle$ initial state discussed above. From the propagators of Appendices and Eqs.~(\ref{conn}) and~(\ref{probability}), one can easily show that $\left|\Psi_+\right\rangle=\frac{1}{\sqrt{2}}(\left|10\right\rangle+\left|01\right\rangle)$ initial state is immune to the applied transverse field; its concurrence remains 1 for every possible value of $\theta$ and $\alpha/\beta$. Note that entanglement of this initial state can be manipulated by using a longitudinal control field as shown in Ref.~\cite{fare}. Note that the interaction Hamiltonian, $\hat{H}_{int}\approx\cos\theta(\hat{\sigma}_1^x+\hat{\sigma}_2^x)+\sin\theta(\hat{\sigma}_1^y+\hat{\sigma}_2^y)$ locally transforms Bell states among themselves for $\theta=0$ and $\theta=\pi/2$. For example $\hat{\sigma}^x_2.\left|\Phi_+\right\rangle=\left|\Psi_+\right\rangle$ and $\hat{\sigma}^x_2.\left|\Phi_-\right\rangle=\left|\Psi_-\right\rangle$ as well as $-i\hat{\sigma}^y_2.\left|\Phi_-\right\rangle=\left|\Psi_+\right\rangle$ and $-i\hat{\sigma}^y_2.\left|\Phi_+\right\rangle=\left|\Psi_-\right\rangle$. It is well known that entanglement is unaffected by local transformations, so one can easily show that the dynamics of $\left|\Phi_+\right\rangle$ at $\theta=0$ is the same as that of $\left|\Phi_-\right\rangle$ at $\theta=\pi/2$ and concurrence remains 1 because the transverse fields transform the two initial states to $\left|\Psi_+\right\rangle=\frac{1}{\sqrt{2}}(\left|10\right\rangle+\left|01\right\rangle)$ which is found to be unaffected by the control sequence as discussed above. Also $\left|\Phi_+\right\rangle$ at $\theta=\pi/2$ and $\left|\Phi_-\right\rangle$ at $\theta=0$ are transformed to $\left|\Psi_-\right\rangle$ and have similar entanglement dynamics; the concurrence for the initial state $\left|\Psi_-\right\rangle$ is independent of $\theta$. At the intermediate values of $\theta$, the local transformations mix different Bell states and it is not straightforward to unentangle them. All of them can be easily checked by using the propagators in the Appendices and the Eqs.~(\ref{conn}) and~(\ref{probability}). The effect of transverse fast pulses on the entanglement dynamics of $\left|10\right\rangle$ or $\left|01\right\rangle$ separable states is qualitatively similar to the results for longitudinal fast pulse case~\cite{fare}, thus we have not covered them.

It was pointed out that the observable non-local time-ordering effects in time are expected to be present if and only if the commutator of the total Hamiltonian~(\ref{totalH}) at different times is nonzero, i.e, $[\hat{H}(t''),\hat{H}(t')]\neq0$~\cite{kaplan,shakov}. By using Eq.~(\ref{totalH}), it is easy to show that  the commutator $[\hat{H}(t''),\hat{H}(t')]$ is equal to
\begin{eqnarray}
\label{com}
[\hat{H}(t''),\hat{H}(t')]=JD\left[\begin{array}{cccc} 0 & a & -a  & 0 \\ -a^* & 0 & 0  & a \\ a^*  & 0 & 0  & -a \\ 0  & -a^* & a^* & 0\end{array}\right],
\end{eqnarray}
where $D=\left((B_2(t')-B_1(t'))-(B_2(t'')-B_1(t''))\right)$ and $a=e^{-i \theta}$. Eq.~(\ref{com}) shows that the commutator vanishes for the cases when $J=0$ and$/$or $B_1(t)=B_2(t)$. By using the numerical solutions of Eq.~(\ref{num}) for the considered initial states and the propagators~(\ref{U^Kpara}),~(\ref{twokickpara}) and~(\ref{tkpara}) in Eqs.~(\ref{conn}) and~(\ref{probability}), in the case of no qubit-qubit interaction~($J=0$) as well as equal magnitude external fields~($\alpha=\beta$), it is easy to show that the concurrence for Bell state is equal to 1, while $\left|11\right\rangle$ state remains separable  at all times and $C(\hat{\rho})$ of both initial states is unaffected from the external kicks and Gaussian pulses. This situation can be also observed from the contour plots for $\alpha=\beta=1$ case. This finding indicates that ability to manipulate the entanglement in the system by using external fields is closely related to the nonlocal time-ordering effects.
\section{Conclusion}
\label{conclusion}
We have investigated the possibility of creation and control of entanglement between two coupled qubits by using time-dependent external magnetic fields in $x$-$y$ plane in the form of delta function kicks and Gaussian pulses for initially pure separable and maximally entangled states. Analytical (for kick sequence) and numerical (for Gaussian pulses) results presented and discussed in the paper indicate a number of interesting phenomena.

Transverse fast pulses can be employed to create long lasting steady entanglement between two coupled qubits initially in a separable state $\left|11\right\rangle$ (or $\left|00\right\rangle$). Furthermore, entanglement of such a state can be finely-tuned by changing the integrated magnetic strength of the external field at qubit positions, while the direction of transverse pulses are found not to effect the entanglement of this particular initial state. Similarly, entanglement of a system initially in a Bell type state is found to be tuned with the same external control parameters; one can even destroy the initial entanglement and recreate this entanglement for this initial state  with carefully designed pulse sequence and system parameters. We have also demonstrated the possibility of obtaining constant long-lived entanglement with desired magnitude by perturbing the qubits with fast pulse or pulses and by adjusting the system parameters to certain values for both initial states.

The time-ordering effect defined by the commutator $[\hat{H}(t_1),\hat{H}(t_2)]$ is found to be important in the manipulation and control of entanglement by fast pulses. It is found that in the case of no time ordering~(i.e., $[\hat{H}(t_1),\hat{H}(t_2)]=0$ which holds for the cases $J=0$ and$/$or $\alpha=\beta$), the transverse fast pulses are found to be ineffective in creation and manipulation of entanglement between qubits initially in any type of pure states.

We have also compared the effect of the external field being longitudinal or transverse and found that with a transverse control one can both create and manipulate entanglement, while the longitudinal fast pulses can only manipulate the entanglement of a state which is already nonzero~\cite{fare}. Moreover, the transverse fast pulses enable one to manipulate, create and control entanglement for a number of initial states, while the manipulation of entanglement with the longitudinal fast pulses was found to be done only with a limited class of initial states~\cite{fare}.

Longitudinal and transverse control fields in the form of kick and Gaussian pulse sequences can be used in a proper sequence to create, control and destroy entanglement from arbitrary initial pure states of two coupled qubits. The same formalism can be applied to initially mixed states by using the time evolution matrix elements provided in Appendices to evolve the density matrix of the system as $\hat{\rho}(t)=\hat{U}(t)\hat{\rho}(0)\hat{U}^{\dagger}(t)$ which might be interesting in analyzing the time evolution of more-general-than entanglement type quantum correlations in mixed states, such as quantum discord~\cite{howz}.

\appendix
\section{Single positive kick}
Here, we consider two qubits whose states are strongly perturbed by external fields which may be expressed as a sudden kick at $t=T$. The time dependent magnetic fields on qubits $1$ and $2$ may be expressed as $B_1(t)=\alpha\delta(t-T)$ and  $B_2(t)=\beta\delta(t-T)$, respectively, where $\alpha$ and $\beta$ are called integrated magnetic strengths. For such a kick the integration over the time is trivial and the time evolution matrix in Eq.~(\ref{U}) becomes as~\cite{fare,shakov}:
\begin{eqnarray}
\label{U^K}
\hat{U}^K(t)=e^{-i\hat{H}_0(t-T)}e^{-i\int_{T-\epsilon}^{T+\epsilon}\hat{H}_{int}(t')dt'}e^{-i\hat{H}_0 T},
\end{eqnarray}
with matrix elements
\begin{eqnarray}
\label{U^Kpara}
U_{11}&=&e^{-iJt}\cos\left(\frac{\alpha}{2}\right)\cos\left(\frac{\beta}{2}\right)=U_{44},\nonumber\\
U_{22}&=&\frac{1}{2}e^{-iJt}\left(e^{4iJt}\cos\left(\frac{\Delta}{2}\right)+\cos\left(\frac{\Omega}{2}\right)\right)=U_{33},\nonumber\\
U_{23}&=&-\frac{1}{2}e^{-iJt}\left(e^{4iJt}\cos\left(\frac{\Delta}{2}\right)-\cos\left(\frac{\Omega}{2}\right)\right)=U_{32},\nonumber\\
U_{12}&=&\frac{1}{2}ie^{-iJt}e^{-i\theta}\left(e^{4iJT}\sin\left(\frac{\Delta}{2}\right)-\sin\left(\frac{\Omega}{2}\right)\right)=e^{-2i\theta}U_{43},\nonumber\\
U_{13}&=&-\frac{1}{2}ie^{-iJt}e^{-i\theta}\left(e^{4iJT}\sin\left(\frac{\Delta}{2}\right)+\sin\left(\frac{\Omega}{2}\right)\right)=e^{-2i\theta}U_{42},\nonumber\\
U_{14}&=&-e^{-iJt}e^{-2i\theta}\sin\left(\frac{\alpha}{2}\right)\sin\left(\frac{\beta}{2}\right)=e^{-4i\theta}U_{41},\nonumber\\
U_{21}&=&-e^{iJ(t-2T)}e^{i\theta}\left(\cos\left(\frac{\beta}{2}\right)\sin\left(\frac{\alpha}{2}\right)\sin(\xi)+i\cos\left(\frac{\alpha}{2}\right)\sin\left(\frac{\beta}{2}\right)\cos(\xi)\right)\nonumber\\
&=&e^{2i\theta}U_{34},\nonumber\\
U_{24}&=&-\frac{1}{2}ie^{-iJt}e^{-i\theta}\left(e^{2i\xi}\sin\left(\frac{\Delta}{2}\right)+\sin\left(\frac{\Omega}{2}\right)\right)=e^{-2i\theta}U_{31},
\end{eqnarray}
where $\Delta=(\alpha-\beta), \Omega=(\alpha+\beta)$ and $\xi=2J(t-T)$. It should be noted that the propagator given by Eq.~(\ref{U^K}) is valid only at times $t>T$. 

\section{Two positive kicks}
The next example is the positive-positive kick sequence applied at times $t=T_1$ and $t=T_2$, namely, $B_1(t)=\alpha(\delta(t-T_1)+\delta(t-T_2))$ and $B_2(t)=\beta(\delta(t-T_1)+\delta(t-T_2))$.  Following the procedure given in Eq.~(\ref{U^K}), one obtains the time evolution matrix at times $t > T_2$ as~\cite{fare,shakov}:
\begin{eqnarray}
\label{kickkick}
\hat{U}^{K}(t)=e^{-i\hat{H}_0(t-T_2)}e^{-i\int_{T_2-\epsilon}^{T_2+\epsilon}\hat{H}_{int}(t')dt'}e^{-i\hat{H}_0(T_2-T_1)}e^{-i\int_{T_1-\epsilon}^{T_1+\epsilon}\hat{H}_{int}(t')dt'}e^{-i\hat{H}_0T_1},\nonumber\\
\end{eqnarray}
with matrix elements
\begin{eqnarray}
\label{twokickpara}
U_{11}&=&\frac{1}{4}e^{-iJt}\left(1+e^{4iJT}\left(\cos(\Delta)-1\right)+\cos(\Delta)+2\cos(\Omega)\right)=U_{44},\nonumber\\
U_{22}&=&\frac{1}{4}e^{-iJt}\left(e^{4iJt}+e^{4iJ(t-T)}\left(\cos(\Delta)-1\right)+e^{4iJt}\cos(\Delta)+2\cos(\Omega)\right)=U_{33},\nonumber\\
U_{23}&=&-\frac{1}{4}e^{-iJt}\left(e^{4iJt}+e^{4iJ(t-T)}\left(\cos(\Delta)-1\right)+e^{4iJt}\cos(\Delta)-2\cos(\Omega)\right)=U_{32},\nonumber\\
U_{12}&=&\frac{1}{2}ie^{-iJt}e^{-i\theta}\left(e^{6iJT}\cos(2JT)\sin(\Delta)-\sin(\Omega)\right)=e^{-2i\theta}U_{43},\nonumber\\
U_{13}&=&-\frac{1}{2}ie^{-iJt}e^{-i\theta}\left(e^{6iJT}\cos(2JT)\sin(\Delta)+\sin(\Omega)\right)=e^{-2i\theta}U_{42},\nonumber\\
U_{14}&=&\frac{1}{4}e^{-iJt}e^{-2i\theta}\left(e^{4iJT}-1-(1+e^{4iJT})\cos(\Delta)+2\cos(\Omega)\right)=e^{-4i\theta}U_{41},\nonumber\\
U_{21}&=&\frac{1}{4}ie^{-iJt}e^{i\theta}\left(e^{4iJ(t-2T)}(1+e^{4iJT})\sin(\Delta)-2\sin(\Omega)\right)=e^{2i\theta}U_{34},\nonumber\\
U_{24}&=&-\frac{1}{4}ie^{-iJt}e^{-i\theta}\left(e^{4iJ(t-2T)}(1+e^{4iJT})\sin(\Delta)+2\sin(\Omega)\right)=e^{-2i\theta}U_{31},
\end{eqnarray}
where $\Delta=(\alpha-\beta)$ and $\Omega=(\alpha+\beta)$. Here, we have assumed equally distanced kicks applied at times $T_1=T$ and $T_2=2T$.

\section{Three positive kicks}
\label{tk}
The final example is the sequence of three positive kicks applied at times
$t=T_1, t=T_2$, and $t=T_3$ namely, $B_1(t)=\displaystyle\sum_{i=1}^3\alpha\delta(t-T_i)$ and $B_2(t)=\displaystyle\sum_{i=1}^3\beta \delta(t-T_i)$.  Following the procedure given in Eq.~(\ref{U^K}), one can obtain the time evolution matrix at times $t > T_3$ as~\cite{fare}:
\begin{eqnarray}
\hat{U}^{K}(t)&=&e^{-i\hat{H}_0(t-T_3)}e^{-i\int_{T_3-\epsilon}^{T_3+\epsilon}\hat{H}_{int}(t')dt'}e^{-i\hat{H}_0(T_3-T_2)}e^{-i\int_{T_2-\epsilon}^{T_2+\epsilon}\hat{H}_{int}(t')dt'}\nonumber\\
&\times&e^{-i\hat{H}_0(T_2-T_1)}e^{-i\int_{T_1-\epsilon}^{T_1+\epsilon}\hat{H}_{int}(t')dt'}e^{-i\hat{H}_0T_1},
\end{eqnarray}
with matrix elements for  $T_1=T, T_2=2T$ and $T_3=3T$,
\begin{eqnarray}
\label{tkpara}
U_{11}&=&\frac{1}{8}e^{-iJt}\left((3-2e^{4iJT}-e^{8iJT})\cos\left(\frac{\Delta}{2}\right)+(1+e^{4iJT})^2\cos\left(\frac{3\Delta}{2}\right)+4\cos\left(\frac{3\Omega}{2}\right)\right)\nonumber\\
&=&U_{44},\nonumber\\
U_{22}&=&\frac{1}{2}e^{-iJt}\left(\cos\left(\frac{3\Omega}{2}\right)+e^{4iJ(t-T)}\cos\left(\frac{\Delta}{2}\right)((1+\cos(4JT))\cos(\Delta)+i\sin(4JT)-1)\right)\nonumber\\
&=&U_{33},\nonumber\\
U_{23}&=&\frac{1}{2}e^{-iJt}\left(\cos\left(\frac{3\Omega}{2}\right)-e^{4iJ(t-T)}\cos\left(\frac{\Delta}{2}\right)((1+\cos(4JT))\cos(\Delta)+i\sin(4JT)-1)\right)\nonumber\\
&=&U_{32},\nonumber\\
U_{12}&=&\frac{1}{2}ie^{-iJt}e^{-i\theta}\left(e^{8iJT}(\cos(4JT)+2\cos(2JT)^2\cos(\Delta))\sin\left(\frac{\Delta}{2}\right)-\sin\left(\frac{3\Omega}{2}\right)\right)\nonumber\\
&=&e^{-2i\theta}U_{43},\nonumber\\
U_{13}&=&-\frac{1}{2}ie^{-iJt}e^{-i\theta}\left(e^{8iJT}(\cos(4JT)+2\cos(2JT)^2\cos(\Delta))\sin\left(\frac{\Delta}{2}\right)+\sin\left(\frac{3\Omega}{2}\right)\right)\nonumber\\
&=&e^{-2i\theta}U_{42},\nonumber\\
U_{14}&=&\frac{1}{8}e^{-iJt}e^{-2i\theta}\left((2e^{4iJT}+e^{8iJT}-3)\cos\left(\frac{\Delta}{2}\right)-(1+e^{4iJT})^2\cos\left(\frac{3\Delta}{2}\right)+4\cos\left(\frac{3\Omega}{2}\right)\right)\nonumber\\
&=&e^{-4i\theta}U_{41},\nonumber\\
U_{21}&=&\frac{1}{2}ie^{-iJt}e^{i\theta}\left(e^{4iJ(t-2T)}(\cos(4JT)+2\cos(2JT)^2\cos(\Delta))\sin\left(\frac{\Delta}{2}\right)-\sin\left(\frac{3\Omega}{2}\right)\right)\nonumber\\
&=&e^{2i\theta}U_{34},\nonumber\\
U_{24}&=&-\frac{1}{2}ie^{-iJt}e^{-i\theta}\left(e^{4iJ(t-2T)}(\cos(4JT)+2\cos(2JT)^2\cos(\Delta))\sin\left(\frac{\Delta}{2}\right)+\sin\left(\frac{3\Omega}{2}\right)\right)\nonumber\\
&=&e^{-2i\theta}U_{31},
\end{eqnarray}
where $\Delta=(\alpha-\beta)$ and $\Omega=(\alpha+\beta)$.

\end{document}